\pdfoutput=1

\documentclass[useAMS,usenatbib]{mn2e}

\usepackage{hyperref}
\usepackage{graphicx}
\usepackage[space]{grffile}
\usepackage{latexsym}
\usepackage{amsfonts,amsmath,amssymb}
\usepackage{url}
\usepackage[utf8]{inputenc}
\usepackage{courier}
\usepackage[T1]{fontenc}
\usepackage{mathptmx}
\usepackage{default}
\usepackage{theorem}
\usepackage{float}
\usepackage{rotating}
\usepackage{lscape}
\usepackage{epsfig}
\usepackage{epstopdf}
\usepackage[usenames,dvipsnames,svgnames,table]{xcolor}
\usepackage{natbib}
\usepackage{threeparttable}
\usepackage[lofdepth,lotdepth]{subfig}
\usepackage[table]{xcolor}
\usepackage{txfonts}

\usepackage{cite}

\newcommand\aap{{\em A\&A}}

\newcommand\aj{{\em AJ}}
\newcommand\apj{{\em ApJ}}

\newcommand\apjs{{\em ApJS}}
\newcommand\araa{{\em ARAA}}

\newcommand\mnras{{\em MNRAS}}

\newcommand\pasp{{\em PASP}}

\newcommand\na{{\em New Astron.}}

\newcommand\physrep{{\em Phys. Rep.}}
\newcommand\memras{{\em MmRAS}}

\title[Polarimetric and Photometric Study of the RCW95 Region]{Optical Polarimetric and Near-Infrared Photometric Study of 
       the RCW95 Galactic H{\sc ii} Region}
\author[J. Vargas-Gonz\'alez et al.]{
J. Vargas-Gonz\'alez$^{1}$\thanks{E-mail: jvargas@dfuls.cl}, 
A. Roman-Lopes$^{1}$, 
F. P. Santos$^{2}$, 
G. A. P. Franco$^{3}$, 
J. F. C. Santos Jr.$^{3}$, 
\newauthor{F. F. S. Maia$^{3}$, 
D. Sanmartim$^{4}$}\\\\ 
$^{1}$Departamento de F\'isica - Universidad de La Serena, Cisternas 1200, La Serena, Chile\\
$^{2}$Physics \& Astronomy Department-CIERA, Northwestern University, Evanston, USA\\
$^{3}$Departamento de F\'isica, ICEx, Universidade Federal de Minas Gerais, Av. Antonio Carlos 6627, 31270-901 Belo Horizonte, 
      MG, Brazil\\
$^{4}$Gemini Observatory, Casilla 603, La Serena, Chile 
}

\begin{document}

\date{Accepted 2016 Month Day. Received 2016 Month Day; in original form 2016 Month Day}

\pagerange{\pageref{firstpage}--\pageref{lastpage}} \pubyear{2016}

\maketitle

\label{firstpage}

\begin{abstract}

We carried out an optical polarimetric study in the direction of the RCW\,95 star forming region in order 
to probe the sky-projected magnetic field structure by using the distribution of linear polarization 
segments which seem to be well aligned with the more extended cloud component. A mean polarization angle of
$\theta=49\fdg8\pm7\fdg7$ was derived. Through the spectral dependence analysis of polarization it was 
possible to obtain the total-to-selective extinction ratio ($R_V$) by fitting the Serkowski function, 
resulting in a mean value of $R_V=2.93\pm0.47$. The foreground polarization component was estimated and 
is in agreement with previous studies in this direction of the Galaxy. Further, near-infrared images from 
Vista Variables in the Via L\'actea (VVV) survey were collected to improve the study of the  stellar population 
associated with the H{\sc ii} region. The Automated Stellar Cluster Analysis  (ASteCA) algorithm was employed to 
derive structural parameters for two clusters in the region, and a set of PAdova and TRieste Stellar Evolution 
Code (PARSEC) isochrones was superimposed on the decontaminated colour-magnitude diagrams (CMDs) to estimate an 
age of about 3 Myr for both clusters. Finally, from the near-infrared photometry study combined with spectra 
obtained with the Ohio State Infrared Imager and Spectrometer (OSIRIS) mounted at the Southern Astrophysics 
Research Telescope (SOAR) we derived the spectral classification of the main ionizing sources in the clusters 
associated with IRAS 15408$-$5356 and IRAS 15412$-$5359, both objects classified as O4 V stars.

\end{abstract}

\begin{keywords}
ISM: individual objects (RCW\,95) -- ISM: magnetic field -- H{\sc ii} regions -- infrared: stars -- open clusters and associations: general
\end{keywords}

\section{Introduction}

Massive star forming regions generally are hosted by highly obscured clouds that in galaxies like the Milky Way, 
are mainly found associated with the spiral arms \citep{b18,b19}. Also, the H{\sc ii} regions generated by the 
massive stars formed there can be identified from the diffuse extended emission produced by the hydrogen 
recombination lines associated with the H$\alpha$ transition in the optical, with the Pa$\beta$ and Br$\gamma$ 
ones in the near-infrared (NIR), as well as by recombination line surveys like the radio one made by \citet{b21}.

The RCW\,95 Galactic star forming region is located at $l=326\fdg7$ and $b=+0\fdg8$ \citep{b24} and presents two bright 
infrared nebula that can be seen in Fig. \ref{rcw95_fig}. \citet{b17} performed a Galactic 5 GHz radio survey in which they 
found a strong continuum source with angular diameter of $3\farcm3$, while \citet{b20}, and \citet{b21}  from radio recombination 
line measurements of the H109$\alpha$ and H110$\alpha$ transitions, found that the main radial velocity component of the associated 
hydrogen recombination line has a mean value of $-44$ km s$^{-1}$, which for a standard Galactic rotation curve corresponds to 
a heliocentric distance of $2.4$ kpc \citep{b30}. A near-infrared study of the stellar content of Galactic H{\sc ii} 
regions performed by \citet{moises2011} gives a range of distances for the RCW95 region from kinematic and non-kinematic (trigonometric 
and K-band spectrophotometric parallaxes) methodologies. A kinematic distance of $2.8\pm0.3$ kpc is given and spectrophotometric 
distances of $1.63\pm0.63$ kpc and $2.02\pm0.78$ kpc result by adopting two extreme interstellar extinction laws from \citet{mathis90} and
\citet{stead09}, respectively. \citet{dutra2003} found two stellar clusters in this Galactic H{\sc ii} region 
located at $l=326\fdg66$, $b=+0\fdg59$ and $l=326\fdg65$, $b=+0\fdg58$ with angular diameters of $2\arcmin$ and $4\arcmin$, 
respectively, estimated visually on the Two Micron All Sky Survey (2MASS, \citet{2mass}) JHKs images. Also, in this region 
\citet{b22,b23} detected two young stellar clusters, both associated with two strong compact mid- to far infrared sources 
cataloged as IRAS15411$-$5352 and IRAS15408$-$5356 . The latter presents at least 136 candidate members in an 
area of about 3 pc$^2$ with about $60\%$ of them presenting some infrared excess emission at $2.2\,\mu$m, characteristic of 
very young stellar clusters \citep{b19}. \citet{Bik2005} and \citet{Bik2006} classified the K-band spectra of 
the brightest and most reddened sources associated to IRAS15408-5356 and IRAS15411-5352, determining spectral types of O5V-O6.5V 
and O8V, respectively.

In this work, one of our goals was to study the interstellar linear polarization component in the direction of the RCW\,95 
Galactic H{\sc ii} Region. The behavior of the magnetic field lines that permeate this star forming region was determined, 
and by applying the empirical Serkowski relation \citep{b16} to the polarimetric data, we were able to compute the total-to-selective 
extinction ratio ($R_V$) mean value in this direction of the Galaxy. Also, another objective was to improve the study of the region's 
stellar population, which was done with point spread function (PSF) photometry on near-infrared VVV images, combined with J-, H- and 
K-band spectroscopic data taken with OSIRIS at SOAR telescope.

\section{Observations and Data Reduction}

\subsection{Optical Polarimetry}

Optical VRI polarimetric data were collected with the 0.9-m telescope of the Cerro Tololo Interamerican Observatory (CTIO, Chile, 
operated by the National Optical Astronomy Observatory--NOAO) during two observing runs in February and April 2010.

The polarimetric module consisted of an arrangement of optical elements placed before the detector along the stellar 
beam path, and composed of a {\it retarder}, an {\it analyzer} and a filter wheel \citep{b25}. The retarder is a half-wave 
plate positioned with an orientation relative to the North Celestial Pole (NCP), which causes a rotation of the polarization 
plane of the incident light beam. It can be rotated in steps of $22\fdg5$, with each $90\degr$ rotation corresponding to a 
polarization modulation cycle. Finally, a Savart plate splits the beam into two orthogonally polarized components (the ordinary 
and extraordinary beams), with each beam passing through spectral filters before finally reaching the CCD detector. 

Image reduction and photometry were done using IRAF\footnote{IRAF is distributed by the National Optical Astronomy Observatory, 
which is operated by the Association of Universities for Research in Astronomy, Inc., under cooperative agreement with the 
National Science Foundation.} tasks. Point-like sources were identified and selected with the DAOFIND task for stars with 
counts 5$\sigma$ above the local background. Magnitude measurements were performed using aperture photometry with 
the PHOT task, for 10 different sized rings around each star.

The polarimetric parameters for all selected sources were computed using a set of specific IRAF tasks that includes FORTRAN 
routines designed for this kind of data (PCCDPACK package, \citealt{b15}). These routines compute the normalized linear 
polarization ($Q$ and $U$ Stokes's parameters) with a least-squares fit solution of the modulation curves (relation of the 
relative intensities between the ordinary and extraordinary beams for each position angle of the half-wave plate), resulting 
in the determination of the degree and angle of linear polarization ($P$ and $\theta$, respectively), with the latter measured 
from the NCP to the east (Eq. \ref{pol_lineal}). In addition, since Eq. (\ref{pol_lineal}) usually generates an overestimation 
of the positive quantity $P$, a correction for this bias was applied as given by Eq. (\ref{P_bias}) \citep{b14}. The $\sigma_P$ 
values adopted were the larger ones between the error from the fitting process and the theoretical error from photon noise \citep{b13}.

\begin{equation}\label{pol_lineal}
 P=\sqrt{Q^2+U^2}\ \ \ \ and \ \ \ \ \theta=\frac{1}{2}\arctan{(U/Q)}
\end{equation}

\begin{equation}\label{P_bias}
P\rightarrow\sqrt{P^2-\sigma_P^2}
\end{equation}

Final calibrations of polarization zero-point angle and degree were made in order to determine possible intrinsic instrumental 
contribution using polarimetric standard stars observed each night  and shown in Table \ref{pol_std_stars}. For the polarization 
degree the intrinsic instrumental polarization contribution is below the typical errors of $\sim0.05\%$. The polarization angle 
corrections from the standard stars correspond to the difference between the polarization angle value from the catalog and those 
from our instrumental measurement ($\Delta\theta=\theta_{cat}-\theta_{inst}$). The errors for the polarization angle were computed 
from $\sigma_{\theta}=28\fdg65(\sigma_{P}/P)$ for $P\gg\sigma_P$ \citep{b12}.

\begin{table}
\caption{Standard stars used for polarimetric calibrations.}
\label{pol_std_stars}
\centering
\renewcommand{\footnoterule}{}
\begin{tabular}{lcccc}
\hline
\hline
\multicolumn{5}{c}{Polarized stars}\\
ID		&Filter	&$P(\%)$	&$\sigma_P(\%)$	&$\theta(\degr)$	\\
\hline
HD111579$^*$	&V	&6.460		&0.014		&103.11		\\  
		&R	&6.210		&0.013		&102.47		\\
		&I	&5.590		&0.017		&102.00		\\
HD110984$^*$	&V	&5.702		&0.007		&91.65		\\
		&I	&5.167		&0.007		&90.82		\\
HD298383$^*$	&V	&5.233		&0.009		&148.61		\\
HD126593$^*$	&R	&4.821		&0.012		&74.81		\\
\multicolumn{5}{c}{Unpolarized stars}\\
ID		&Filter	&$Q(\%)$	&\multicolumn{2}{c}{$U(\%)$}	\\
\hline
HD150474$^{**}$	&V	&0.001		&\multicolumn{2}{c}{-0.007}	\\
HD126593$^{***}$	&V	&-0.05		&\multicolumn{2}{c}{-0.03}	\\
\hline
\hline
\end{tabular}
\begin{tablenotes}
 \item $*$ \citet{turn90}.
 \item $**$ \citet{gil2003}.
 \item $***$ \citet{fossati2007}.
\end{tablenotes}

\end{table}

\begin{figure}
  \includegraphics[width=0.5\textwidth]{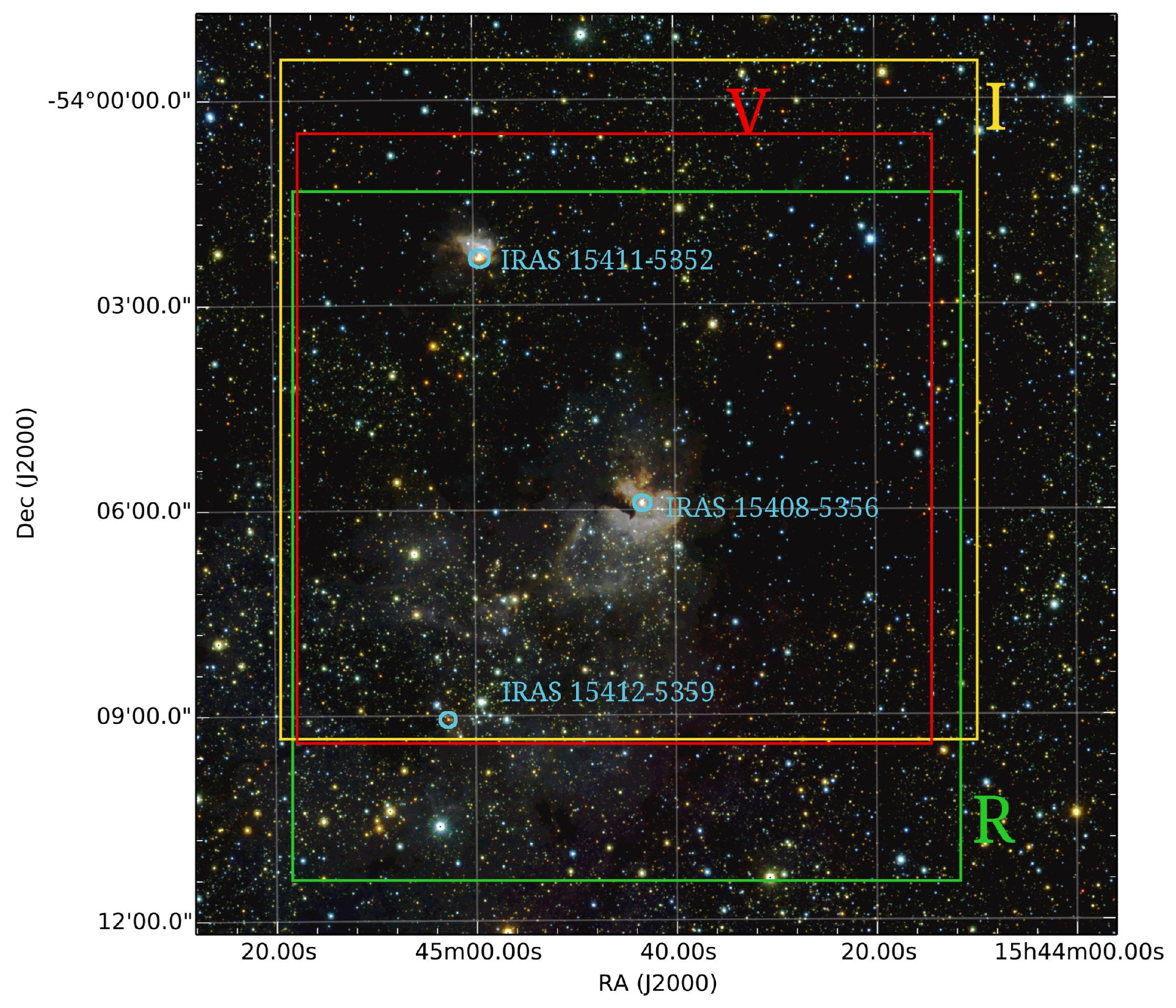}
  \caption{Combined RGB image from the J, H and Ks VVV images centered on RCW\,95 H{\sc ii} region. The coloured boxes 
           indicate the polarization survey areas related to different spectral bands. The V optical band covers an area 
           of $9\farcm0\times8\farcm6$ (red box), the R optical band covers an area of $9\farcm6\times10\farcm0$ (green box) and the I 
           optical band covers an area of $10'\times10'$ (yellow box). Three IRAS sources in the field are indicated in light 
           blue.}
\label{rcw95_fig}           
\end{figure}

The polarization survey areas are shown in Fig. \ref{rcw95_fig} in different colour boxes related to each spectral 
band. The V optical band covers an area of $9\farcm0\times8\farcm6$ (red box), the R optical band covers an area 
of $9\farcm6\times10\farcm0$ (green box) and the I optical band covers an area of $10\arcmin \times10\arcmin$ 
(yellow box). A sample of the polarimetric survey is shown in Table \ref{pol_cat}.

\begin{table}
\caption{The V, R and I Optical polarimetric sample.}
\label{pol_cat}
         {\scriptsize
   	\begin{tabular}{llllllll}
  	\hline
  	\hline
         $\alpha_{2000} ({\rm \ ^h\ ^m\ ^s} )$ & $\delta_{2000} ( \degr \ \arcmin \ \arcsec )$& ID  &Filter&$P(\%)$&$\sigma_P(\%)$&$\theta(\degr)$&$\sigma_{\theta}(\degr)$\\
  	\hline  	
         15:44:30.09                    &-54:03:04.9                       & 193 &  V   &5.698  & 1.080        &      51.22      &   5.42  \\
                                                &                                         &        &  R   &7.388  & 0.513        &      59.24      &   1.98  \\
                                                &                                         &        &  I    &5.401  & 0.697        &      56.01      &   3.69  \\                                             
        15:44:34.22                     &-54:03:06.4                       & 194 &  V   &6.181  & 0.270        &      55.82      &   1.25  \\
                                                &                                         &        &  R   &6.170  & 0.185        &      56.64      &   0.85  \\
                                                &                                         &        &  I    &5.017  & 0.275        &      56.01      &   1.57  \\                                               
        15:44:32.92                     &-54:03:09.1                       & 195 &  V   &6.386  & 1.023        &      56.62      &   4.59  \\
                                                &                                         &        &  R   &5.717  & 0.412        &      61.14      &   2.06  \\
                                                &                                         &        &  I    &6.379  & 0.780        &      56.81      &   3.50  \\
        15:45:11.06                     &-54:03:08.9                       & 196 &  V   &3.147  & 3.226        &      61.22      &   29.36 \\
                                                &                                         &        &  R   &3.082  & 0.443        &      53.44      &   4.12  \\
                                                &                                         &        &  I    &1.739  & 0.420        &      60.11      &   6.91  \\               
        15:45:18.10                     &-54:03:10.7                       & 197 &  V   &  ---     &   ---            &     ---            &  ---    \\
                                                &                                         &        &  R   &  ---     &   ---            &     ---            &  ---    \\
                                                &                                         &        &  I    &8.575  & 4.233        &      45.41      &   14.14 \\                                              
        15:44:41.78                     &-54:03:11.6                       & 198 &  V   &  ---     &   ---            &     ---            &  ---    \\
                                                &                                         &        &  R   &4.156  & 0.708        &      60.24      &   4.88  \\
                                                &                                         &        &  I    &6.004  & 0.867        &      57.61      &   4.13  \\
         \hline
         \hline
  	\end{tabular}  
 \begin{tablenotes}
 \item Note: Polarization angle and degree values marked with $-$ correspond to objects without data in this specific spectral 
             band.
 \end{tablenotes}
         }
 \end{table}

\subsection{Near-infrared Photometry}\label{vvv_phot}

The near-infrared photometric image data in the J ($1.25\,\mu$m), H ($1.65\,\mu$m) and Ks ($2.15\,\mu$m) bands were 
retrieved from the ESO Public Survey VVV designed to scan the inner Milky Way \citep{b26} with the VISTA InfraRed 
CAMera (VIRCAM) at the 4.1-m Telescope at Cerro Paranal Observatory, in Chile \citep{b27}. VIRCAM provides observations 
with a field of view of $1.48\times1.11$ deg$^2$ and a spatial resolution of $0\farcs34$ per pixel. In this 
study we have selected the Tile d098 in the J, H and Ks bands of the VVV survey covering  the field of RCW\,95 H{\sc ii} 
region around $\alpha_{J2000}=15^{\rm h}44^{\rm m}5\fs5$ and $\delta_{J2000}=-53\degr54^{\rm m}24^{\rm s}$, which comprises 
the area of the polarimetric survey in this work.

\begin{table}
\centering
\caption{List of PSF radii and PSF fitting radii for each filter applied for the PSF photometry.}
\label{psf_radii}
\resizebox{8.7cm}{!}{
\begin{tabular}{@{}lcccccc@{}}
\hline
			&\multicolumn{2}{c}{J}&\multicolumn{2}{c}{H}&\multicolumn{2}{c}{Ks}\\
			& (pixels)&(\arcsec)  & (pixels) & (\arcsec)& (pixels)&(\arcsec)   \\
\hline
 PSF Radius 		& 11      &3.74 	   & 11       &3.74      &    11   &3.74        \\
 PSF Fitting Radius& 2.5     &0.85       & 3.9      &1.326     &   3.7   &1.258       \\
\hline
\end{tabular}
}
\end{table}

\begin{figure}
  \includegraphics[width=0.1565\textwidth]{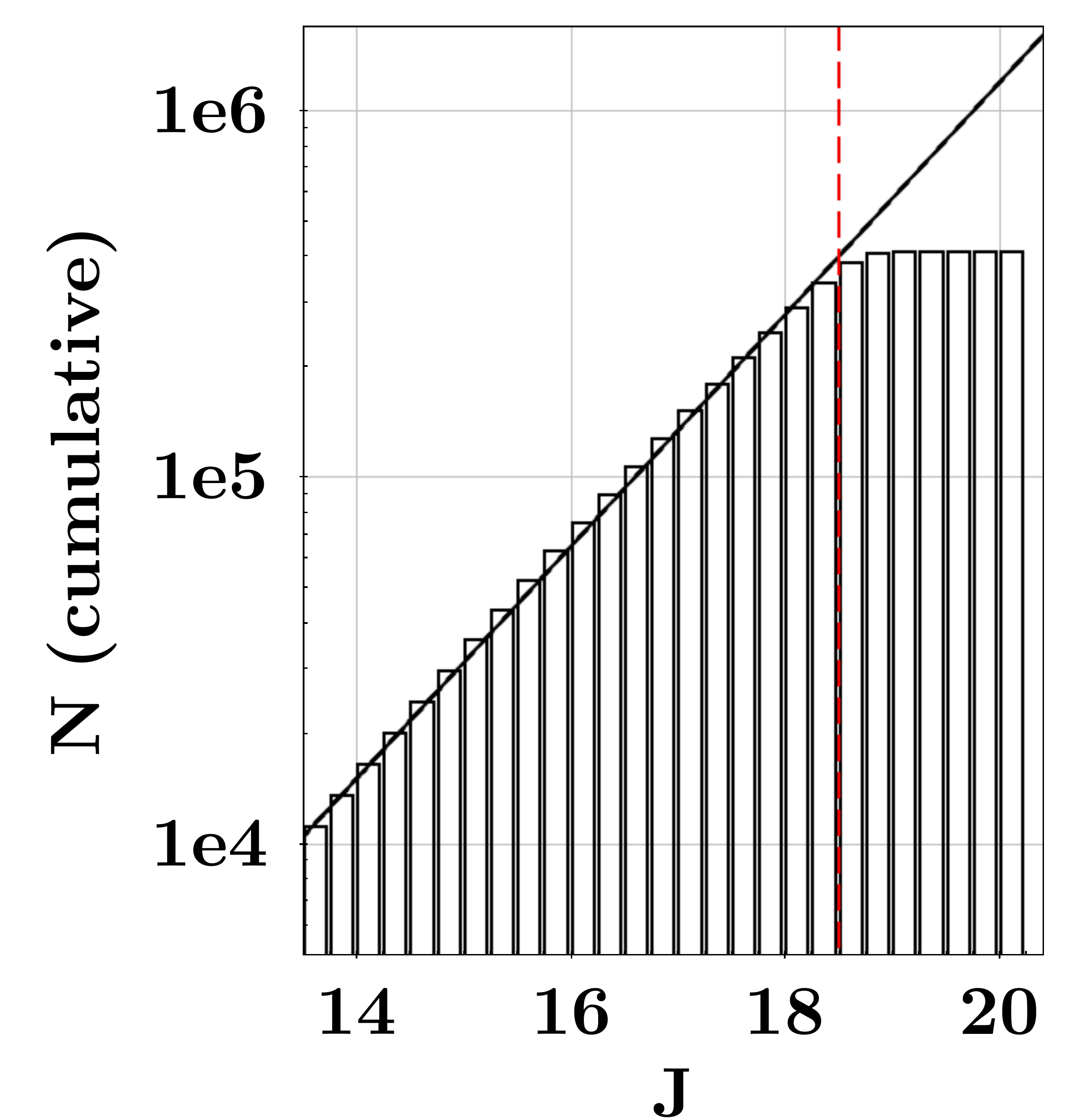}
  \includegraphics[width=0.1565\textwidth]{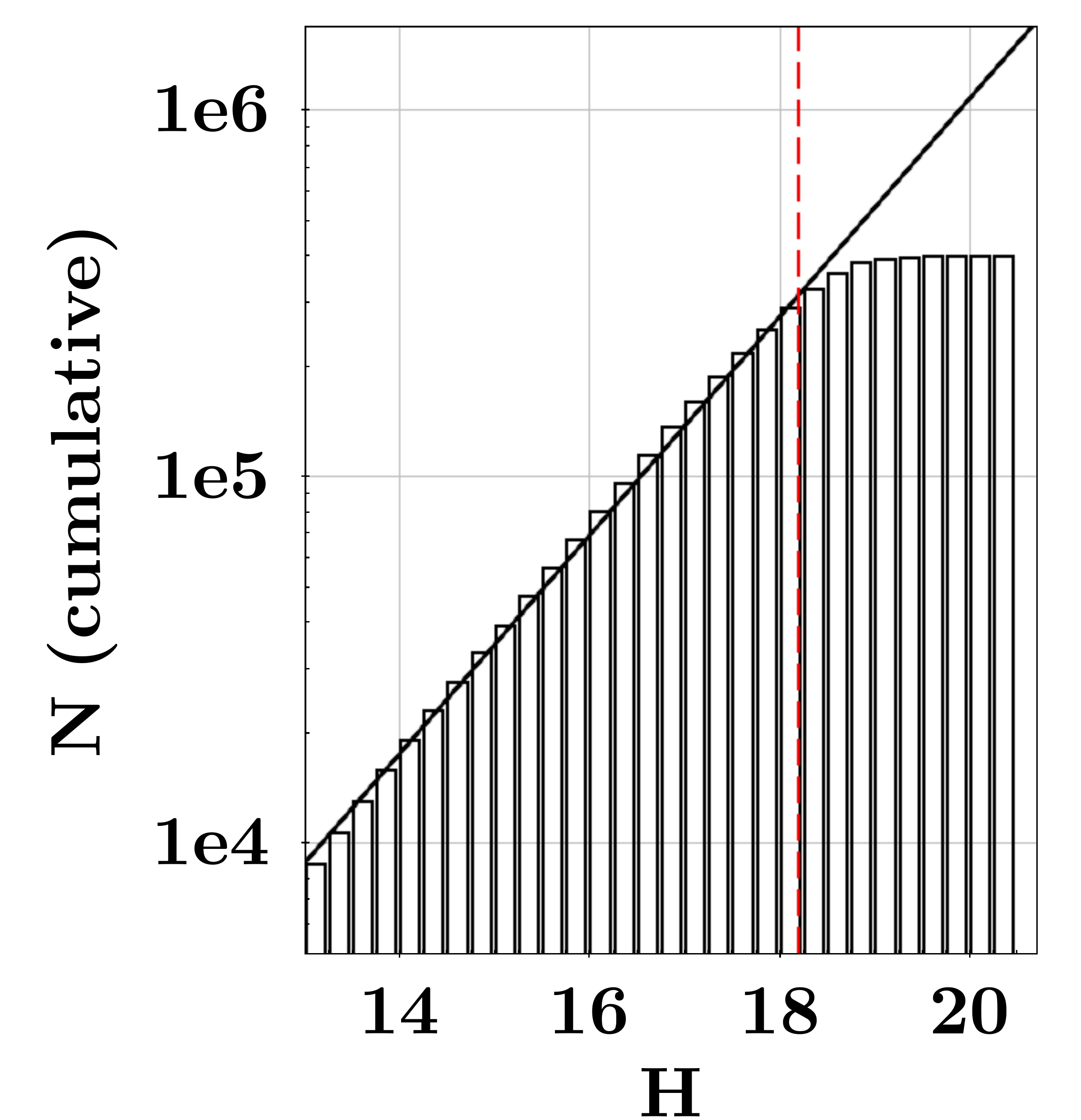}
  \includegraphics[width=0.1565\textwidth]{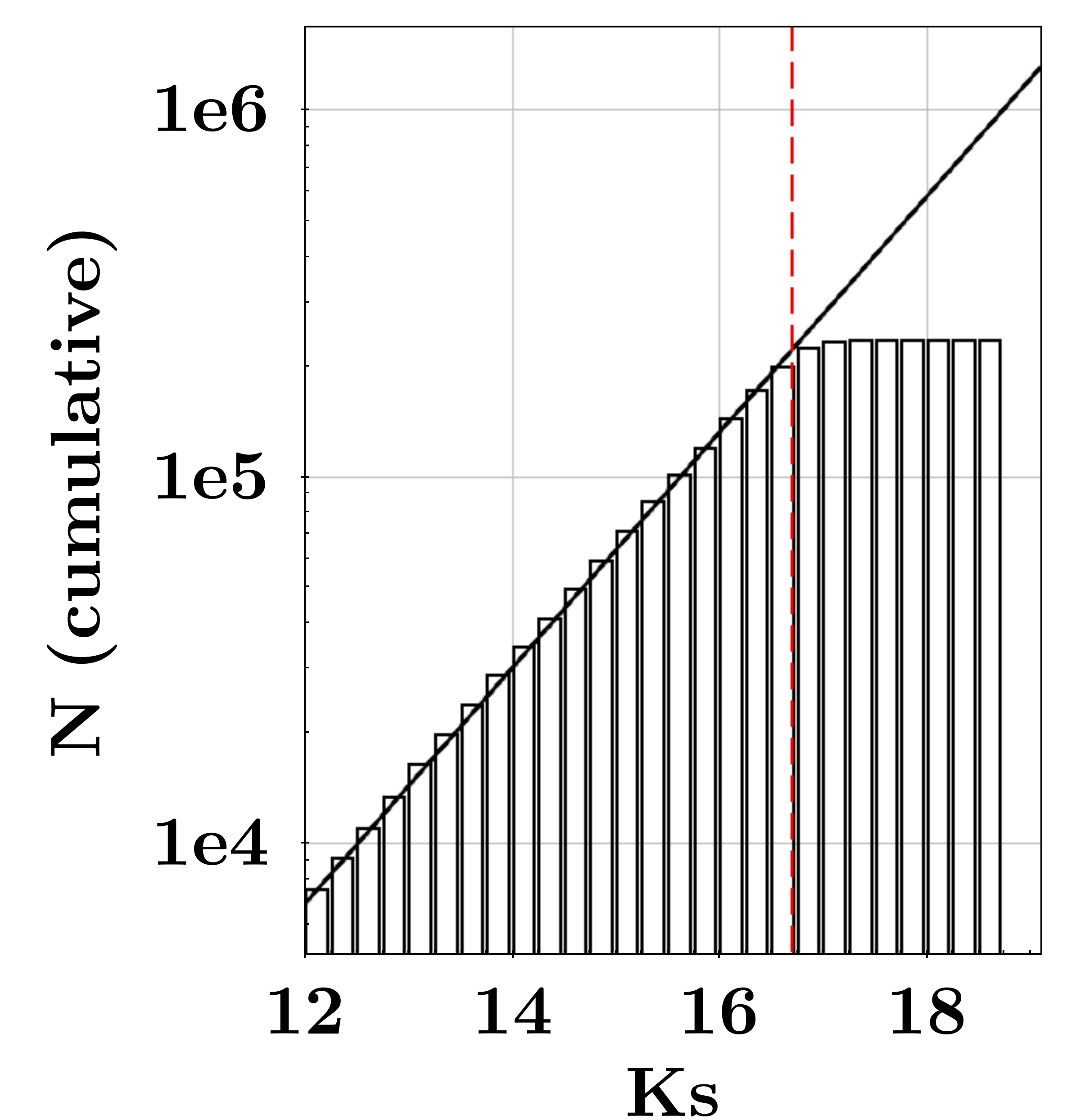}
  \caption{Cumulative source counts as function of magnitude used to  derive the 
                        completeness limit (cl) for each filter indicated by the vertical dashed 
                        lines.}
  \label{completeness}
\end{figure}

Because of crowding within the RCW\,95 H{\sc ii} region, PSF photometry was performed using IRAF tasks 
from the DAOPHOT package \citep{b28} for stars peaked 10$\sigma$ above the local background.
The best fitting for the PSF was obtained from 'penny2' function. The PSF radius and PSF 
fitting radii used for each filter are indicated in Table \ref{psf_radii}. The PSF fitting and stellar 
subtraction process were run several times in order to reveal and obtain the magnitudes for the very close 
faint companions in the most crowded areas. The completeness limits values for J, H and Ks bands 
are $18.5$, $18.2$ and $16.7$ mag, respectively, and were estimated from the distribution of cumulative star 
counts as a function of magnitude, as shown in Fig. \ref{completeness}, where a linear relation was fit to 
determine the point at which the number of detected sources decreases in the distribution. The above completeness 
limits represent an improvement of about $1.1$, $2.0$ and $3.1$ magnitudes in the J, H and Ks bands, respectively, 
when compared to the earlier survey (aperture photometry) by \citet{b22}. The completeness resulting from 
this procedure are similar to artificial star tests for uncrowed fields as shown by \citet{m16}. In order to recover photometric information 
from saturated stars in the VVV images they were replaced with 2MASS magnitudes transformed to the VVV photometric 
system according to \citet{soto2013}. A sample of the photometric results is shown in Table \ref{phot_cat}.

\begin{table*}
\centering
\caption{NIR photometric sample from VVV data. Columns 1 and 2 are the equatorial coordinates 
         {\it $(\alpha$, $\delta)_{J2000}$}, The following columns are the magnitudes together
         with each corresponding uncertainties.}
\label{phot_cat}
\resizebox{13cm}{!} {
\begin{threeparttable} 	
        {\small
  	\begin{tabular}{llllllll}
 	\hline
 	\hline
        $\alpha_{2000} (\ ^{\rm h}\ ^{\rm m}\ ^{\rm s})$ & $\delta_{2000} ( \degr \ \arcmin \ \arcsec )$&J&$\Delta J$&H&$\Delta H$&Ks&$\Delta Ks$\\
 	\hline  	
        15:45:04.13  &	$-$54:03:35.4  &	 14.396	&  0.020  &  10.256	&  0.024  &  8.037   &  0.018\\
        15:45:09.82  & 	$-$54:10:37.9  &	 14.353	&  0.023  &  10.277	&  0.024  &  8.106   &  0.021\\
        15:43:55.29  &	$-$53:48:49.2  &	 13.531	&  0.026  &  11.077	&  0.022  &  9.932   &  0.019\\
        15:43:12.50  &	$-$54:23:27.9  &	 13.965	&  0.030  &  11.345	&  0.030  &  10.043  &  0.021\\
        15:42:27.68  &	$-$53:54:36.7  &	 17.890	&  0.027  &  13.514	&  0.023  &  11.480  &  0.025\\
        15:45:20.00  &	$-$54:00:16.2  &	 17.734	&  0.031  &  13.943	&  0.020  &  12.066  &  0.026\\
        15:42:22.11  &	$-$54:20:20.7  &	 14.938	&  0.021  &  13.087	&  0.021  &  12.197  &  0.024\\
        15:43:26.96  &	$-$54:27:48.9  &	 14.904	&  0.022  &  13.129	&  0.058  &  12.204  &  0.023\\
        15:39:59.02  &	$-$54:05:54.6  &	 15.815	&  0.032  &  13.924	&  0.025  &  13.045  &  0.025\\    
        15:43:35.90  &	$-$54:17:29.4  &	 18.222	&  0.031  &  15.370	&  0.030  &  14.019  &  0.020\\ 
 	\hline
        \hline
 	\end{tabular}  
}
\end{threeparttable}
}
\end{table*}

\subsection{Near-infrared Spectroscopy}

\begin{table}
\caption{Summary of the SOAR/OSIRIS dataset used in this work.}
\label{spect_summary}
\centering
\renewcommand{\footnoterule}{}
\begin{tabular}{cc}
\hline
   Date  & 03/04/2007\\
         & 04/04/2015\\
   Telescope  & SOAR\\
   Instrument & OSIRIS\\
   Mode  & XD\\
   Camera & f/3\\
   Slit  & 1" x 27"\\
   Resolution  & 1000\\
   Coverage ($\mu$m) & 1.25-2.35\\
   Seeing (")  & 0.8-1.1\\
   Targets & IRS1 / IRS2 / HD303308\\
\hline
\end{tabular}
\end{table}

NIR spectroscopic observations for the two brightest sources selected from our photometric study were performed with the 
OSIRIS at the SOAR telescope. The J-, H- and K-band spectroscopic data were acquired on April 4th 2015, with 
the night presenting good seeing conditions. We also used the NIR spectra of HD\,303308 as template and the spectroscopic 
data for this object was acquired by Alexandre Roman-Lopes on March 4th 2007. Table \ref{spect_summary} summarises the NIR 
observations.

The spectroscopic dataset was reduced following standard NIR reduction procedures that are detailed in \citet{b31}, and 
shortly explained here. The dispersed images were subtracted for each pair of images taken at the two positions along 
the slit. Next the resultant images were divided by a master normalized flat, and for each processed frame, the J-, H- 
and K-band spectra were extracted using the IRAF task APALL, with subsequent wavelength calibration being performed using 
the IRAF tasks IDENTIFY/DISPCOR applied to a set of OH sky line spectra (each with about 30-35 sky lines in the range 
12400 - 23000 \AA). The typical error (1$\sigma$) for this calibration process is estimated as $\sim$12 \AA\ 
which corresponds to half of the mean FWHM of the OH lines in the mentioned spectral range. Telluric atmospheric corrections 
were done using J-, H- and K-band spectra of A type stars obtained before and after the target observations. The photospheric 
absorption lines present in the high signal-to-noise telluric spectra, were subtracted from a careful fitting (through the use 
of Voigt and Lorentz profiles) to the hydrogen absorption lines and respective adjacent continuum. Finally, the individual J-, 
H- and K-band spectra were combined by averaging (using the IRAF task SCOMBINE) with the mean signal-to noise 
ratio of the resulting spectra being well over 100.

\section{Results and Analysis}

\subsection{Optical Polarimetry}

\subsubsection{Map of the distribution of Polarization Segments}

\begin{figure*}
  \includegraphics[width=0.6\textwidth]{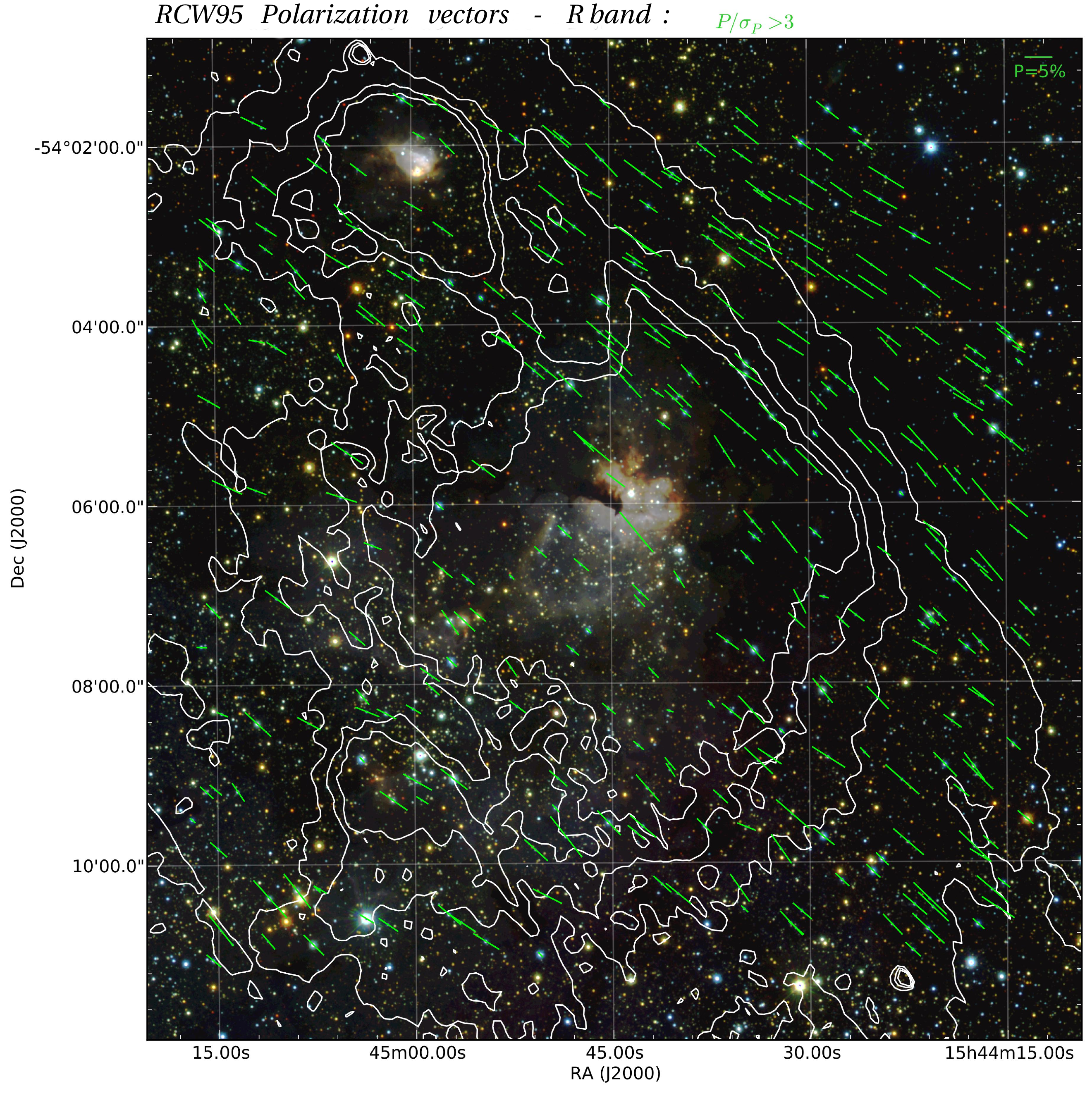}
  \caption{Spatial distribution map of the R band polarization segments in the direction of the RCW\,95 H{\sc ii} region 
          (area$\sim10'\times10'$) with north to the top and east to the left. The sizes of each segment are proportional to 
          the polarization degree (an equivalent to a $5\%$ polarization degree segment is indicated in the right corner of 
          the image). Each segment's orientation is measured from the NCP. Just $P/\sigma_P>3$ data are shown. The countour 
          lines correspond to the {\it Spitzer}-MIPS $70\mu$m over a RGB composition from VVV data in the J, H and Ks 
          bands.}
\label{R_pol_map}
\end{figure*}

\begin{figure*}
  \includegraphics[width=0.72\textwidth]{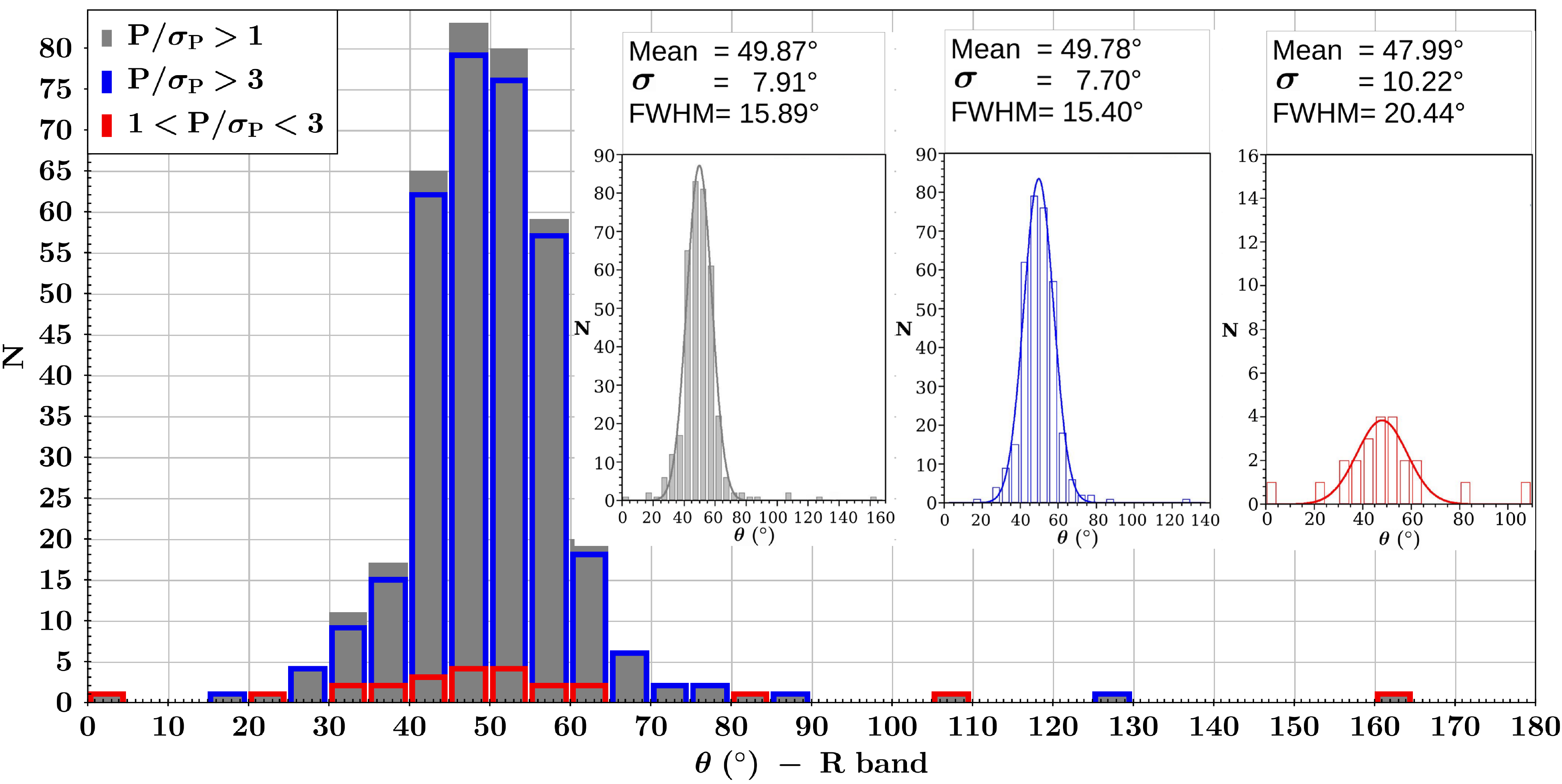}
  \caption{R band polarization angle histograms. Each colour represent the quality data where grey correspond to all data 
          with $P/\sigma_P>1$, blue for data with $P/\sigma_P>3$ and red for data with 
          $1<P/\sigma_P<3$. The mean polarization angle from a Gaussian fit to each distribution results in 
          $49\fdg87$, $49\fdg78$ and $47\fdg99$ for the grey, blue and red distributions, respectively. There is an 
          increase in the dispersion with lower values of $P/\sigma_P$ (from $\sigma_\theta=7\fdg70$ for the blue 
          distribution up to $\sigma_\theta=10\fdg22$ for the red one).}
\label{R_angle_histo}
\end{figure*}

\begin{figure}
  \includegraphics[width=0.45\textwidth]{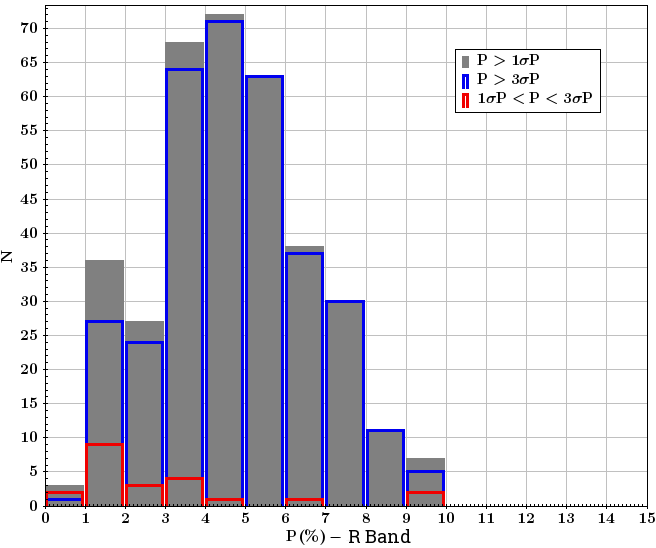}
  \caption{R band polarization degree histograms. Coloured distributions correspond to the quality data criteria used 
           in Fig. \ref{R_angle_histo}. The polarization degree for $P/\sigma_P>3$ data range between 
           $P=0\ -\ 10\%$ with a higher frequencies between $P=3\ -\ 6\%$ ($\sim 60\%$ of the blue distribution sources 
           lies within this range).}
\label{R_degree_histo}
\end{figure}

From the optical polarimetric survey it was possible to derive the spatial distribution of the linear polarization segments 
in the direction of RCW\,95. The resulting R-band map can be seen in Fig. \ref{R_pol_map}. There they are represented by 
green segments of line that are plotted on a RGB composite image (centred on the main cluster) made from the J (blue), H 
(green) and Ks (red) bands of VVV images, with each segment representing the degree of polarization and polarization angle 
($P$ and $\theta$, respectively) of each star with good photometric measurements in the observed field. The size of each 
segment is proportional to the polarization degree (the scale reference for a $5\%$ polarization degree is indicated at 
the top-right corner of the map), and the orientation representing the polarization angle measured from NCP (equatorial 
reference). As a complement and in order to identify a possible correlation between the magnetic field component projected 
onto the plane of the sky, and the more extended dust component of the molecular cloud, we also constructed a contour map 
that was derived from a {\it Spitzer-MIPS} $70\,\mu$m image taken from the NASA/IPAC Infrared Science Archive 
(IRSA\footnote{http://irsa.ipac.caltech.edu/Missions/spitzer.html}).

The polarization angle maps appear similar in all spectral bands, with the mean values being compatible considering the 
uncertainties ($\langle\theta_V\rangle=51\fdg2\pm7\fdg6$, $\langle\theta_R\rangle=49\fdg8\pm7\fdg7$ and 
$\langle\theta_I\rangle=49\fdg4\pm8\fdg9$). As can be seen from the polarization map distribution obtained for the R 
band data, Fig. \ref{R_pol_map}, our sample contains a high number of sources with polarization values obeying the 
condition $P/\sigma_P >3$ ($82\%$, $88\%$ and $78\%$ for the V, R and I bands, respectively). In the analysis of the 
results, we choose to be conservative by excluding all sources presenting $P/\sigma_P$ values not satisfying the mentioned 
condition. Indeed, from an inspection of the results obtained for the associated objects, it was concluded that most of 
them correspond to bad quality data measurements mainly resulting from the presence of bad pixels, cosmic rays on the 
detector and/or sources close to the borders of the images, where the photometric sensitivity drops considerably. 
However, in order to enrich the discussion sometimes the low quality polarimetric data corresponding to sources with 
$P/\sigma_P <3$ were also used in order to complement the associated analysis.

As can be seen from the background NIR composite image (Fig. \ref{R_pol_map}), the interstellar extinction in this direction of 
the Galaxy is high and highly variable, with the two star clusters clearly seen to the center and to the north-east, apparently 
associated with the most optically obscured parts of the region. Also, the spatial distribution of the two clusters seems well 
correlated with the $70\,\mu$m contours with the extended cloud surrounding them in a somewhat elongated geometrical configuration. 
This is important because it could indicate that in this case the expansion of the H{\sc ii} regions tends to follow the magnetic 
field (B) lines perhaps suggesting a correlation between the mean B orientation and the Galactic plane. In this sense, we can see 
that the overall polarization segments distribution seems to be well aligned with the more extended cloud component, with a mean 
polarization angle of $49\fdg8$ as inferred from the histograms of polarization angles in Fig. \ref{R_angle_histo}. In this figure 
we highlight the three distributions related to the $P/\sigma_P$ quality data criteria stated above. There the blue one corresponds 
to sources with $P/\sigma_P >3$ ("good quality data"). A Gaussian fit was applied to each distribution resulting in a mean 
polarization angle for the R band of $\theta_R=49\fdg8\pm7\fdg7$. There is a correlation in the mean value for all quality data 
but with higher dispersion toward lower values of $P/\sigma_P$ up to $\sigma_\theta=10\fdg2$ for the red distribution. In addition, 
in order to evaluate some correlation between the mean B orientation and the Galactic plane it is possible to convert the polarization 
angles from equatorial to Galactic coordinates according to \citet{stephens2011}. The Galactic polarization angle for the R band is 
then $\theta_G=87\fdg0$ which means that the mean B orientation shows a good correlation with the Galactic plane (the position angle 
of Galactic plane is $90\degr$).

The histograms for the observed values of the polarization degree are shown in Fig. \ref{R_degree_histo}. There we 
also can see those corresponding to the three distributions described earlier as shown in Fig. \ref{R_angle_histo}. 
The polarization degree values for the best quality data ranges between $P=0\ -\ 10\%$ with the higher frequencies 
occurring for values between $P=3\ -\ 6\%$ (with $\sim 60\%$ of the blue distribution sources lying within this 
range). The red distribution does not show any apparent preferential range of polarization degree values, however 
they correspond to the $11\%$ of the R band entire sample. 

\subsubsection{Wavelength dependence of polarization degree and the observed total-to-selective extinction 
               ratio}\label{serkowski}
               
\begin{figure*}
  \includegraphics[width=0.281\textwidth]{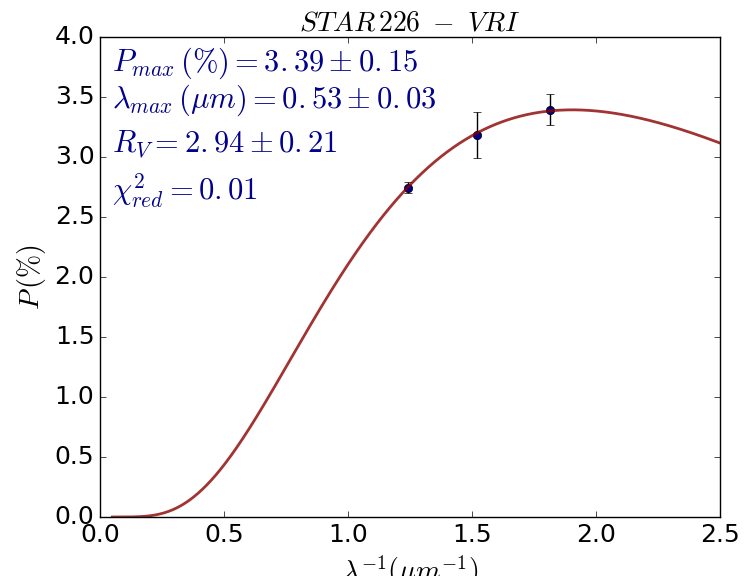}
  \includegraphics[width=0.271\textwidth]{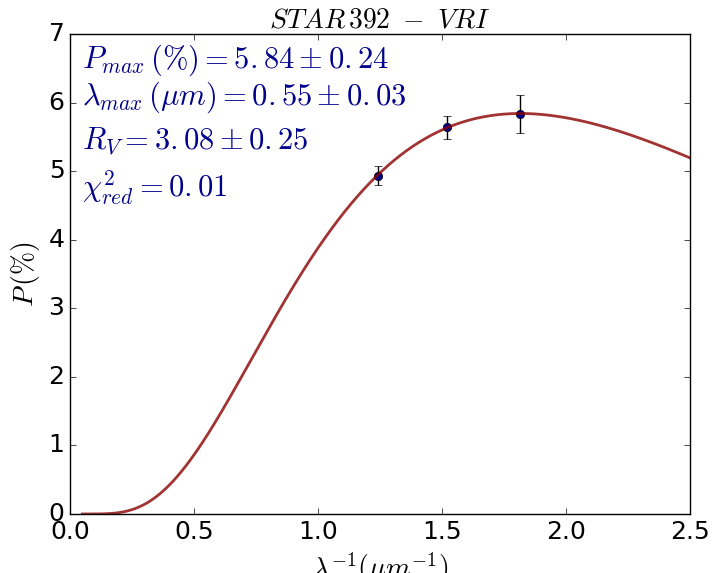}
  \includegraphics[width=0.55\textwidth]{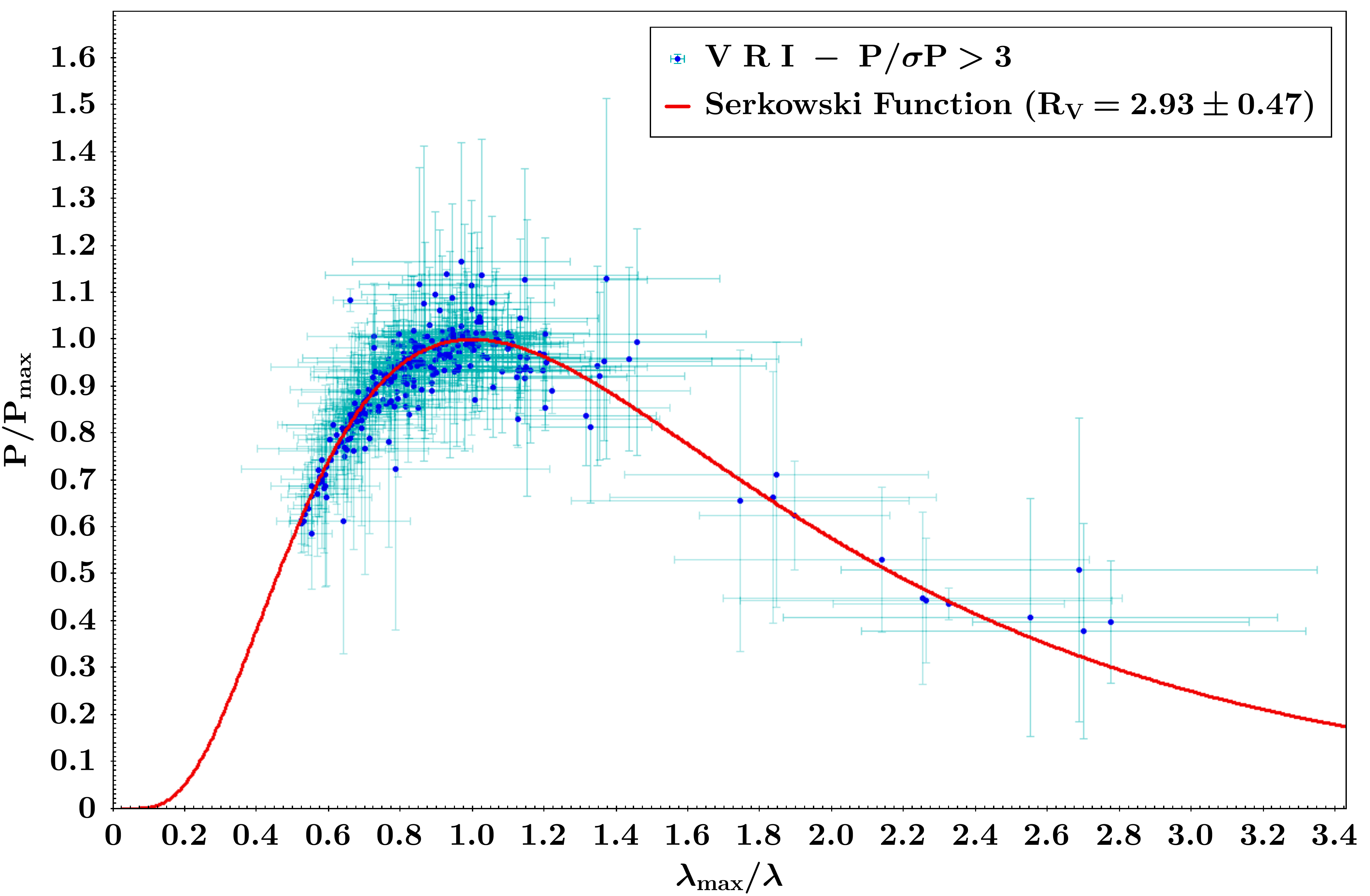}
\caption{({\it Top}) Serkowski fitting in a $P(\%)$ vs. $\lambda^{-1}(\mu m^{-1})$ diagram for two sorces whose 
         polarimetric parameters obey the criteria $P/\sigma_P>3$ and belong to the set of curves with best fits 
         (lower values of the reduced chi-square $\chi^2_{red}$). Each fitted value of $P_{max}$, $\lambda_{max}$ 
         and $R_V$ is listed at the top. ({\it Bottom}) Serkowski relation in a $P/P_{max}\times\lambda_{max}/\lambda$ 
         diagram for all sources obeying $P/\sigma_P>3$ as well as $\chi^2_{red}<2$ (blue dots with their error bars 
         in light-blue). The red line indicates the Serkowski fit for $R_V=2.93\pm0.47$.}
\label{serkowski_all}
\end{figure*}
               
The V, R and I band polarimetric survey in principle allow us to study the wavelength dependence of the 
polarization degree, which in turn can be related to specific properties of the interstellar reddening law 
in the direction of RCW\,95 H{\sc ii} region. This wavelength dependence can be described by the Serkowski 
empirical relation \citep{b16}:

\begin{equation}\label{serkowski_law}
 P_{\lambda} = P_{max}\exp{\Bigg[ -K\ \ln^2\Bigg ( \frac{\lambda_{max}}{\lambda} \Bigg ) \Bigg]}
\end{equation}
where $P_{\lambda}$ and $P_{max}$ are the polarization degree at a wavelength $\lambda$ and the maximum 
polarization level computed from the fitting, respectively, with $\lambda_{max}$ corresponding to the 
wavelength where such maximum is reached. The $K$ parameter determines the width of the peak in the 
Serkowski curve, and in general it is assumed as $K=1.15$ for the optical spectral range \citep{b16}.

Based on the polarimetric dataset derived from the V, R and I bands observations ($197$ sources with good 
measurements), we selected all sources presenting $P/\sigma_P>3$ in all three bands ($153$ sources, $\sim78\%$), 
and constructed the $P(\%)$ vs. $\lambda^{-1}(\mu{\rm m}^{-1})$ diagrams shown in the top panel of 
Fig. \ref{serkowski_all}. The fitting process to each star was performed using the EMCEE\footnote{Python 
implementation of the affine-invariant ensemble sampler for Markov chain Monte Carlo (MCMC) proposed by 
\citet{mcmc}.} code adapted for the Serkowski relation applied to the individual values of $P_{max}$ and 
$\lambda_{max}$ for each source in our sample \citep{emcee,seron}. The $P(\%)$ 
vs. $\lambda^{-1}(\mu{\rm m}^{-1})$ diagrams seen in the top panel of Fig. \ref{serkowski_all} 
show examples of the Serkowski relation (red lines) applied for two sources in our sample, and by 
using the mentioned parameters of Eq. (\ref{serkowski_law}), we computed the associated $R_V$ values 
with the Eq. (\ref{r_v}) \citep{b16,b29}: 

\begin{equation}\label{r_v}
 R_V=(5.6\pm 0.3)\lambda_{max}
\end{equation}
with the derived parameters being indicated in the top-left corner of each 
$P(\%)\times\lambda^{-1}(\mu{\rm m}^{-1})$ diagram in the Fig. \ref{serkowski_all}.

In order to calculate the mean value of $R_V$ in the direction of RCW\,95, we applied the fitting process 
only to the sources presenting $\chi^2_{red} < 2$ from the Serkowski relation 
fitting. From the entire sample, 106 sources meet the mentioned conditions resulting in a mean total-to-selective 
extinction ratio value (weighted average) $R_V=2.93\pm0.47$ which, given its uncertainty, can be compared to 
the standard value of $R_V=3.09\pm0.03$ for Galactic interstellar grains \citep{rieke1985}.

In the bottom panel of Fig. \ref{serkowski_all} we present the Serkowski relation plotted over the 
106 sources (blue dots with error bars in light-blue) used to compute the mean $R_V$ value in the 
direction of RCW\,95. The uncertainties on the derived polarimetric parameters increase to the right 
side of the plot just toward lower $\lambda$ values, as expected because the heavy extinction in 
general contributes to decrease the signal-to-noise ratio values toward bluer colours, mainly for 
the stars in the direction of RCW\,95.

\subsubsection{Foreground R-band polarization component}\label{foreground_pol_component}

 \begin{figure}
   \centering
   \includegraphics[width=0.35\textwidth]{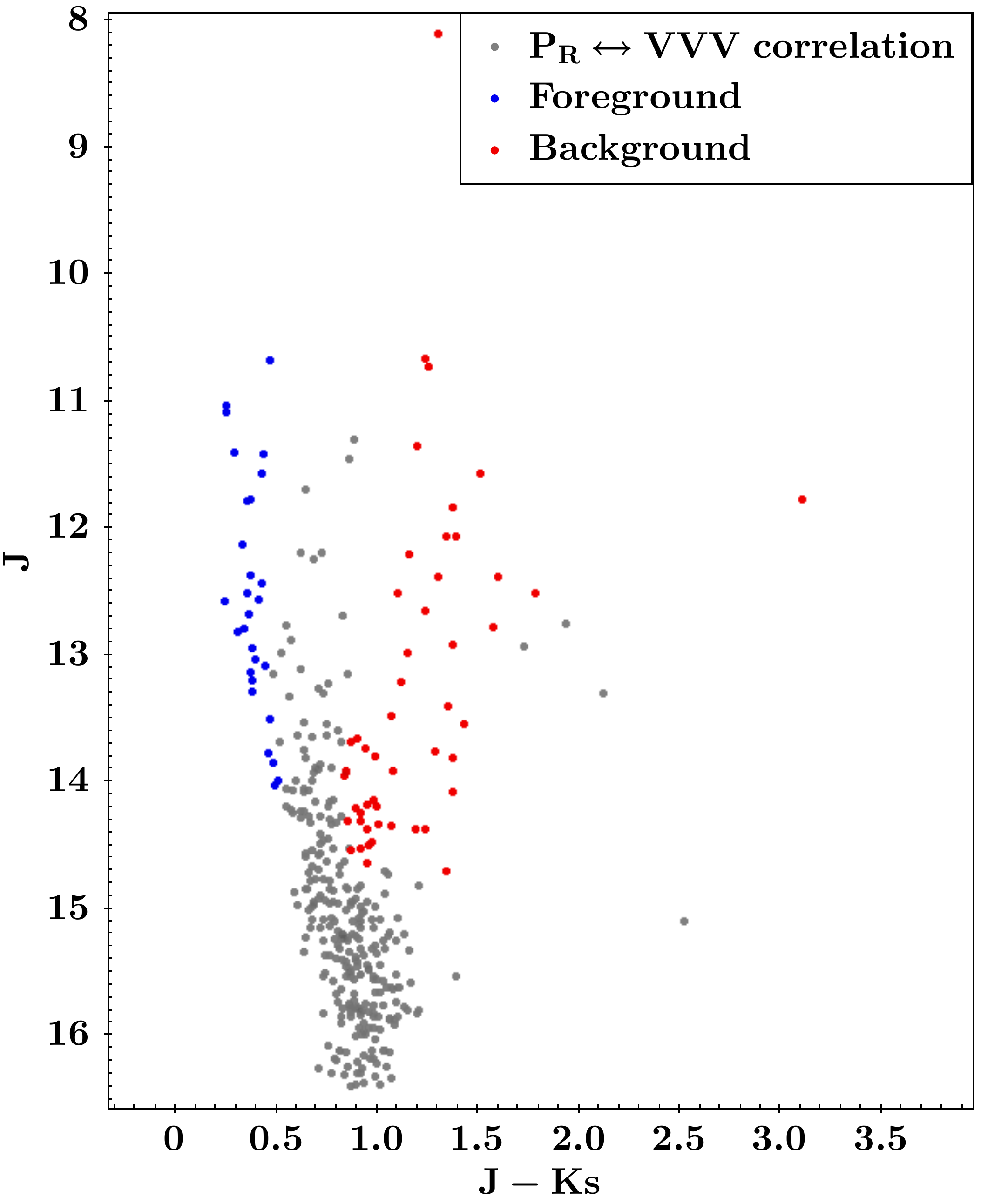}
   \caption{Colour-magnitude diagrams from the correlation between R polarimetric and JHKs photometric 
            catalogs. There are shown in red dots sources with high extinction along the reddening band that 
            most probably are giant field Galactic sources behind the star forming region (background 
            stars). Blue dots correspond to foreground sources with lower interstellar absorption values 
            around the unreddened main-sequence locus.}
\label{fore_back_separation}
\end{figure}

\begin{figure}
   \centering
   \includegraphics[width=0.4\textwidth]{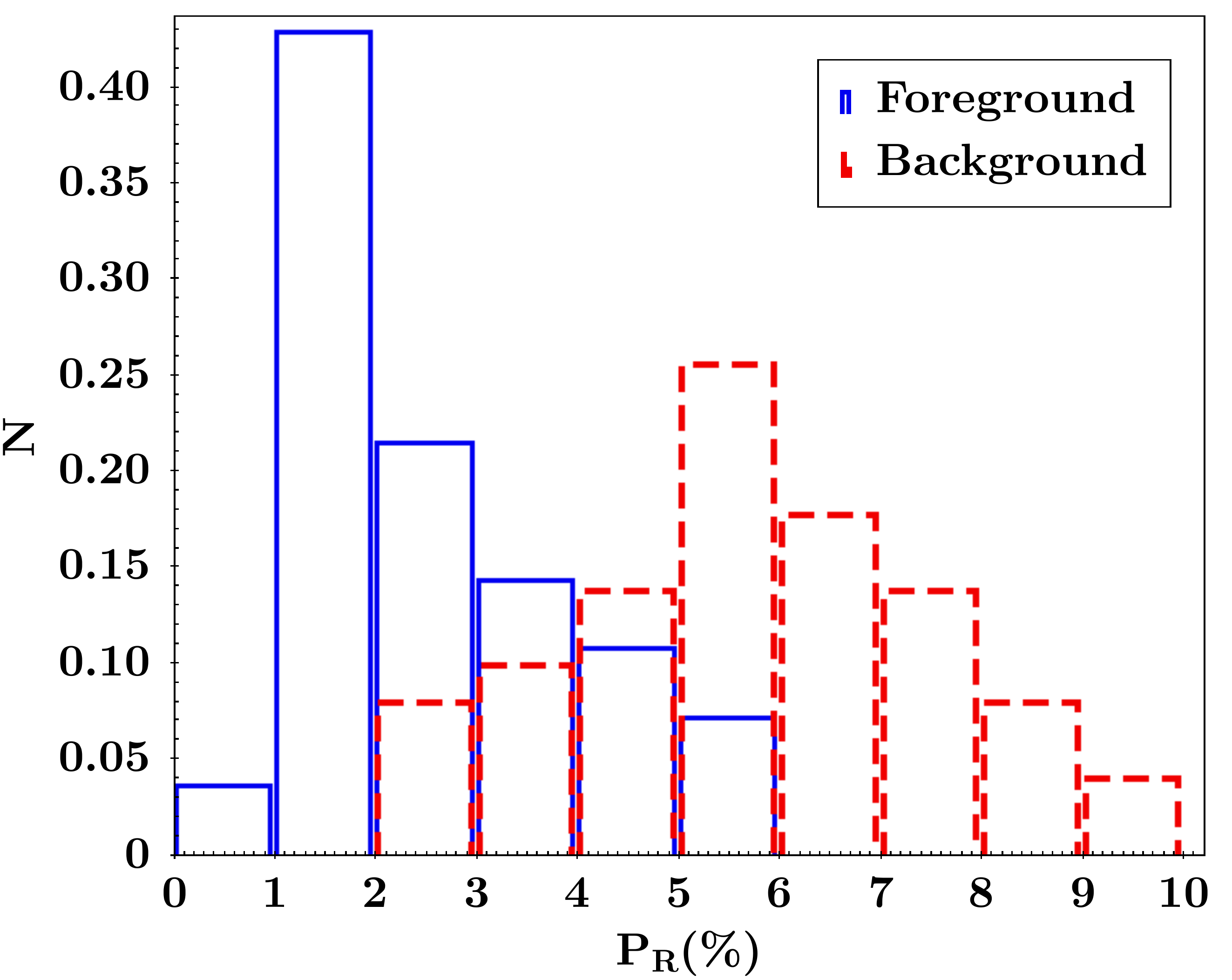}
   \caption{R-band polarization degree ($P_R$) histograms. Sources identified as foreground population are 
            indicated by blue dots while red ones are those identified as background population, according 
            to the analysis from Fig. \ref{fore_back_separation}.}
\label{histo_P_norm}
\end{figure}

\begin{figure}
   \centering
   \includegraphics[width=0.28\textwidth]{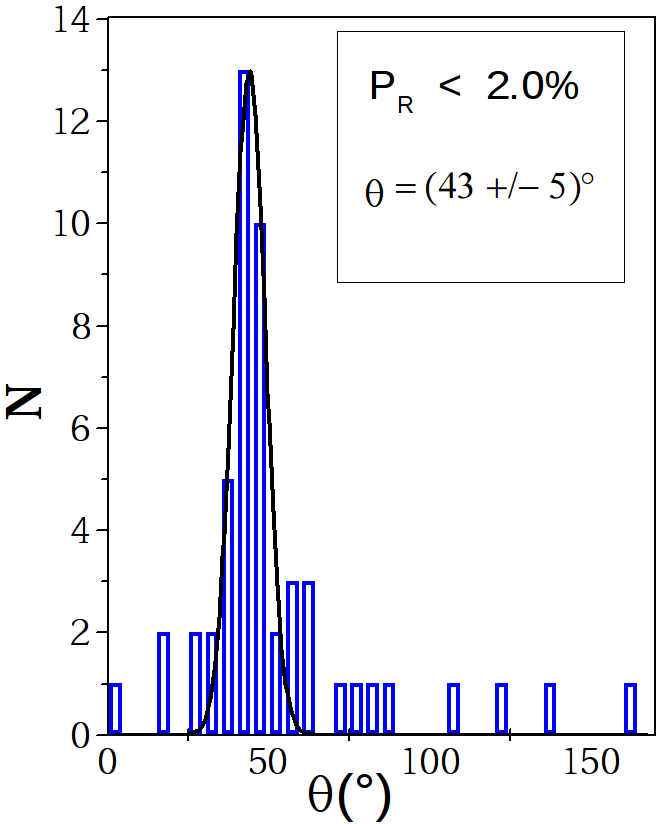}
   \includegraphics[width=0.28\textwidth]{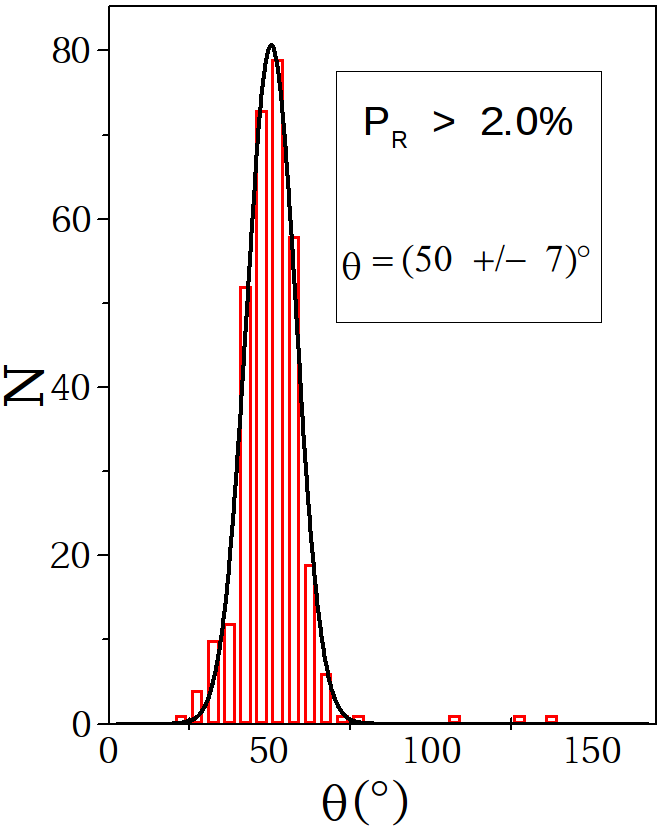}
   \caption{Polarization angle distribution for sources presenting $P_R<2.0\%$ {\it (top panel)} and those 
            presenting $P_R>2.0\%$ {\it (bottom panel)} from the R-band polarimetric catalog.}
\label{histo_P_2}
\end{figure}

For an assumed heliocentric distance of 2.4\,kpc \citep{b30}, it is expected that at least part of 
the polarization levels measured for the stars in our sample are possibly affected by a line of sight 
foreground interstellar component. In this sense, we proceed to estimate this foreground 
contribution by using the NIR photometric colours of 
the probable foreground and background sources, taking into account the observed connection with the 
associated polarimetric parameters. The NIR photometric colours taken from the results presented in Sect. 
\S\ref{vvv_phot_results} were correlated with our polarimetric sample to identify the sources 
corresponding to the main-sequence and red-giant and/or super-giant Galactic disk population, as can be seen 
in Fig. \ref{fore_back_separation}. There the blue distribution corresponds to the low-reddened sources in the 
CMD, which in turn are associated with the main-sequence disk stars. On the other hand, the probable giant field 
sources (e.g. those mainly located behind the star forming region) correspond to the red dots. Both 
the blue and the red sample were manually selected from the mentioned locus in the CMD avoiding to select sources 
located in ambiguous position in the diagram. The observed polarization degree distributions of this two groups of 
objects are shown in Fig. \ref{histo_P_norm}, which is colour coded in the same way. We can see that the polarization 
degree levels for the foreground sources peaks around $P_R=2.0\%$, with no background sources showing polarization 
degree lower than this value in the red distribution, which we assume as the probable mean value of the foreground 
component of the interstellar polarization value in this direction of the Galaxy. 

We also analyzed the polarization angle distribution for stars with $P_R<2.0\%$ and $P_R>2.0\%$, and by 
Gaussian fittings we estimated the mean polarization angle associated with each distribution. The results 
are shown in Fig. \ref{histo_P_2}. The angles $\theta=43\degr$ for the foreground component and $\theta=50\degr$ 
for the background component show that there is no significant difference between the polarization angle for 
the two source groups resulting in a mean polarization angle of $\theta=46\fdg5\pm8\fdg5$. From a search in 
the associated literature we found that \citet{mathewson1971} derived values $P$=2.19\% and $\theta=51\degr$ 
for HD\,141318, a star in the vinicity of the RCW\,95, both values well in line with our estimates for the 
mean polarization angle and foreground polarization component in this part of the Galaxy.

The Stokes parameters associated with the foreground component can be estimated using the Eq. (\ref{pol_lineal}) as follow: 

\begin{equation}
\label{stokes_foreground}
\begin{split}
 Q_f&=P\cos{2\theta}\\
 U_f&=P\sin{2\theta}
\end{split}
\end{equation}
where $Q_f$ and $U_f$ are the Stokes parameters associated with the foreground component with 
$P$ and $\theta$ estimated above, resulting in values of $Q_f=0.137$ and $U_f=1.955$, respectively. 

\subsection{VVV near-infrared photometric study}\label{vvv_phot_results}

In the following sections we describe the study of the stellar population in the direction of 
the RCW\,95 H{\sc ii} region. The stellar clusters studied are those associated with the infrared 
sources IRAS 15408$-$5356 and IRAS 15412$-$5359, hereafter identified as 15408 and 15412, respectively.

\subsubsection{Cluster parameter determinations}\label{param_struct}

\begin{figure}
  \includegraphics[width=0.41\textwidth]{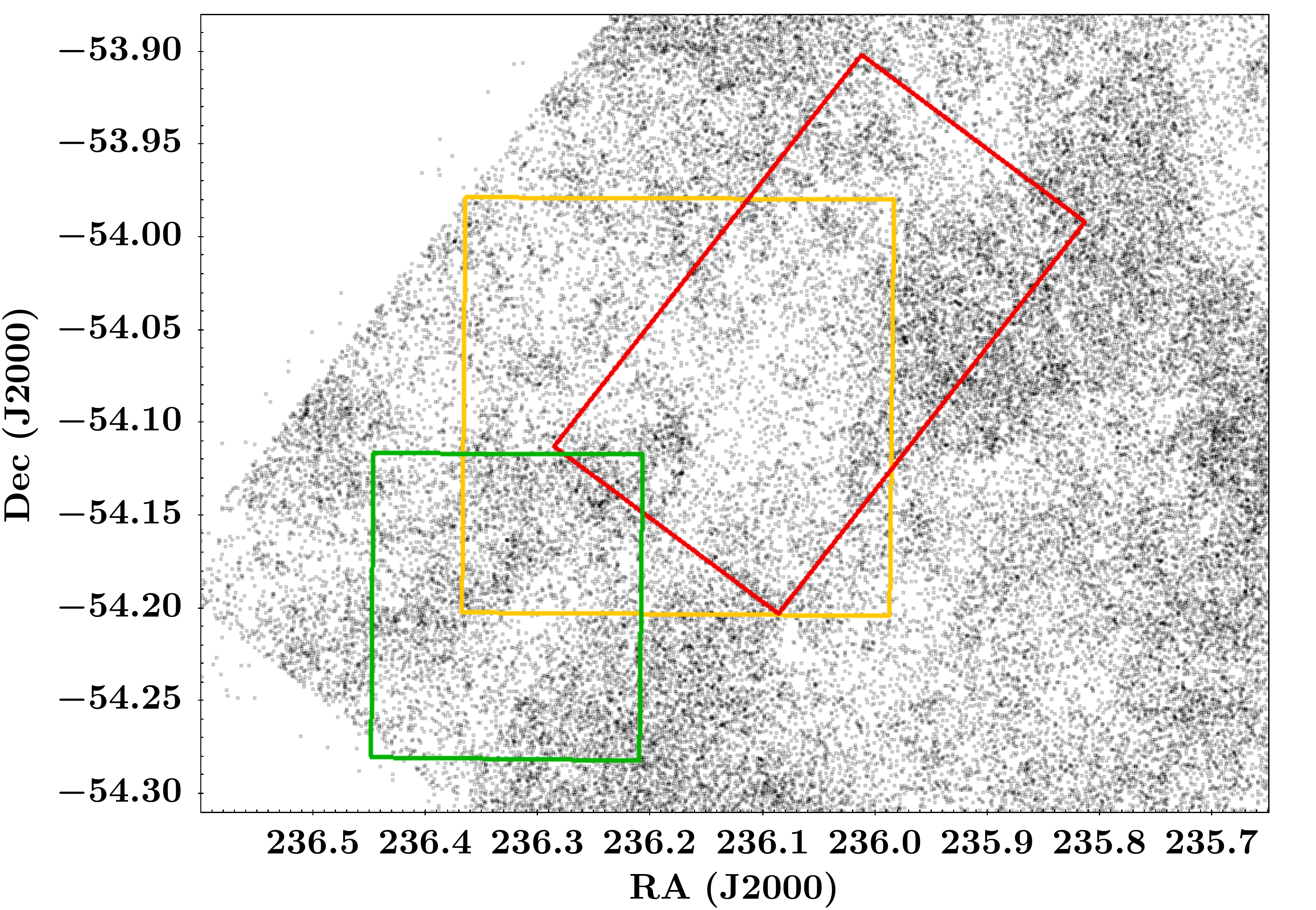}
  \includegraphics[width=0.395\textwidth]{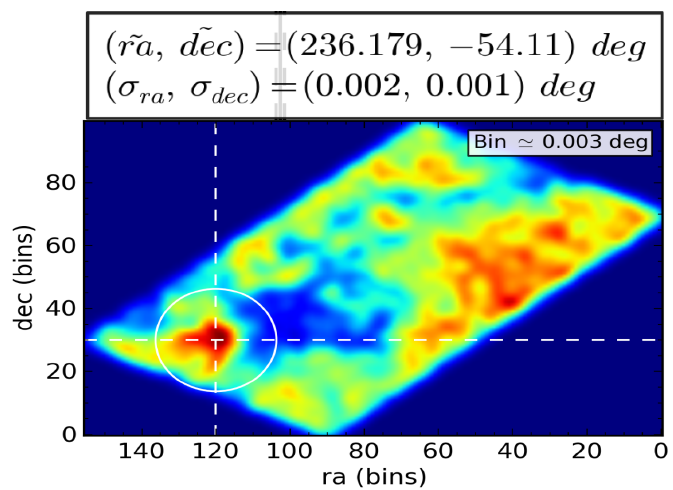}
  \includegraphics[width=0.4\textwidth]{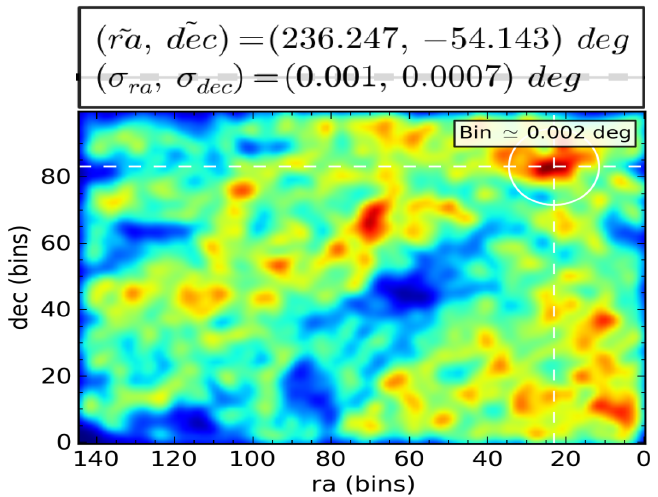}
  \caption{({\it} top) Areas of the center determination analysis were red and green correspond to the areas of the clusters 15408 and 15412, 
           respectively. The middle and bottom panles show the center determination via a two-dimesional KDE. In the header of each panel are 
           shown the center coordinates values together with its uncertainties.}
  \label{KDE_center}
\end{figure}

\begin{figure}
   \centering
        \includegraphics[width=0.47\textwidth]{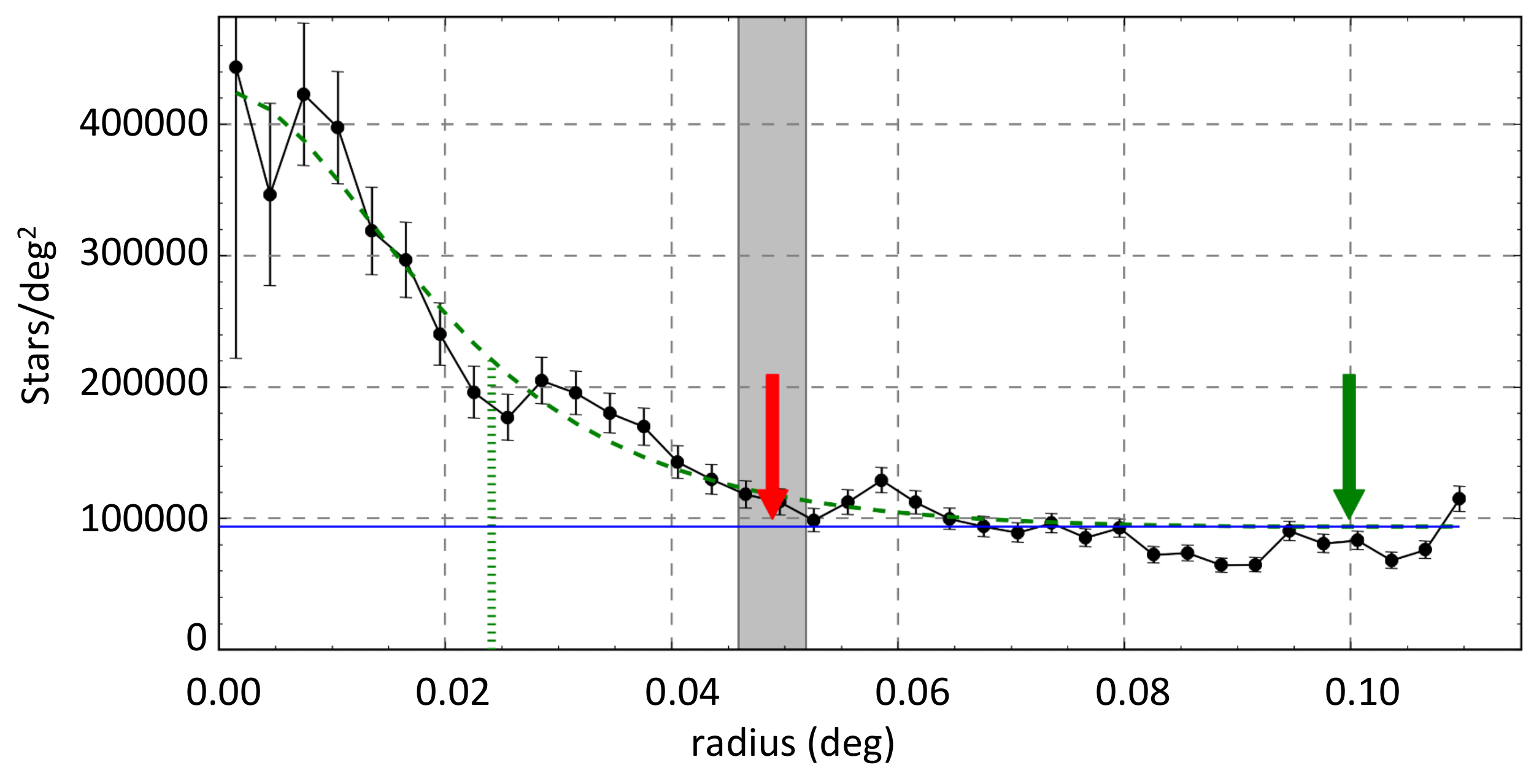}
        \includegraphics[width=0.47\textwidth]{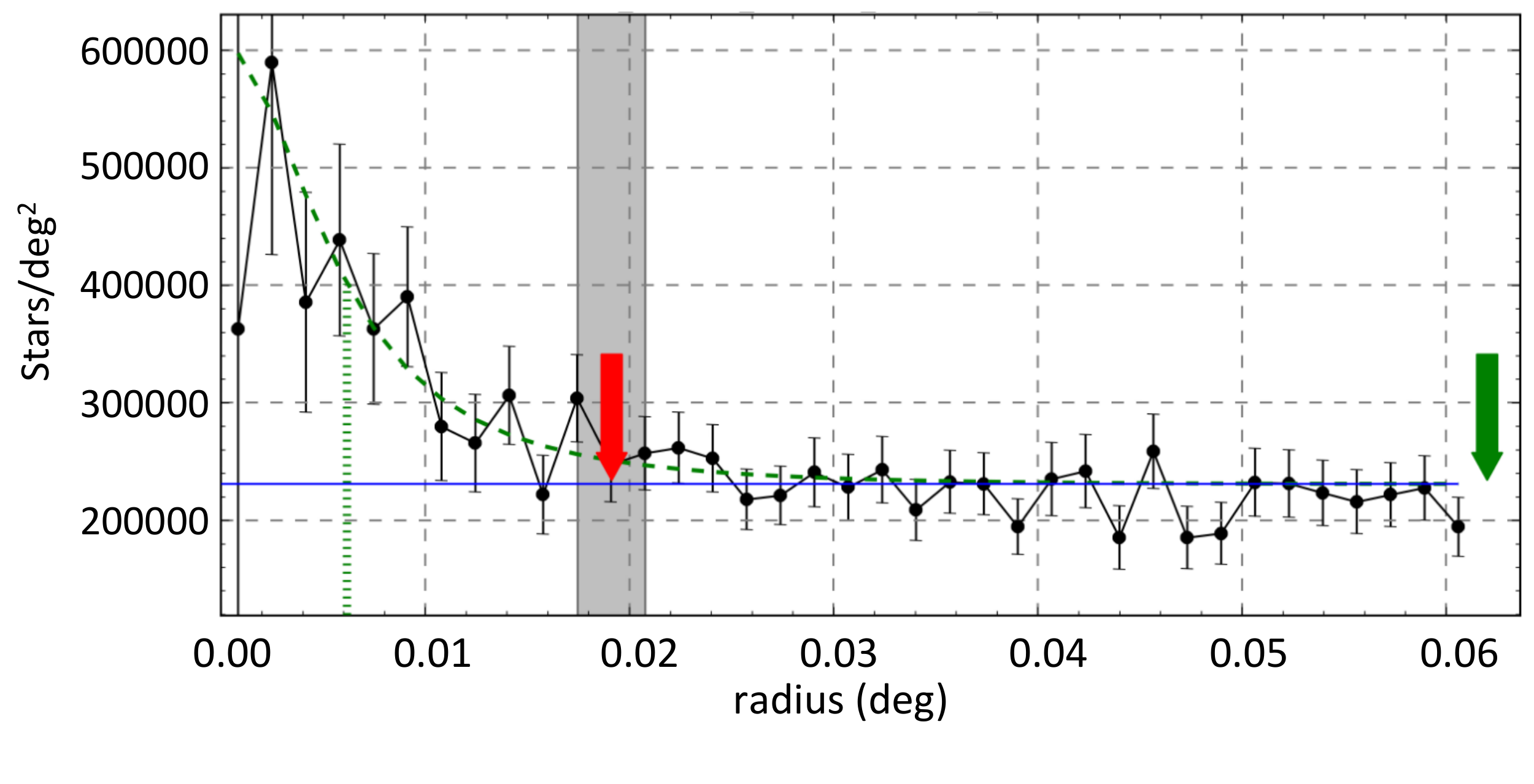}
	\caption{Radial density profile of the regions of clusters 15408 and 15412. The black dots correspond to the RDP point from 
	         the center of each cluster. The 3-parameter King profile is represented by the dashed green line, the vertical
	         green line indicates the cluster core radius and the tidal radius is represented by the green arrow. The horizontal 
	         blue line indicates the $\sigma_f$ value and the cluster radius with its uncertainties are indicated by the
	         red arrow and the grey region, respectively.}
	\label{RDP}
\end{figure}

The initial reference coordinates for each cluster are those of the IRAS sources, which are $\alpha_{\rm J2000}=15^{\rm h}44^{\rm m}42\fs8$ 
and $\delta_{J2000}=-54\degr05\arcmin56\arcsec$ for IRAS 15408$-$5356 and $\alpha_{J2000}=15^{\rm h}45^{\rm m}03\farcs4$ and 
$\delta_{J2000}=-54\degr09\arcmin08\arcsec$ for IRAS 15412$-$5359. In order to determine the best values of the centre for each 
cluster, we made use of the Automated Stellar Cluster Analysis (ASteCA) algorithm \citep{perren2015}. The latter allowed us to 
compute the central coordinates of the cluster as the point with maximum spatial density value, by fitting a two-dimensional 
Gaussian kernel density estimator (KDE) to the density map of each cluster. In Fig. \ref{KDE_center} we show the results of the 
2D Gaussian KDE fitting for 15408 ({\it middle}) and 15412 ({\it bottom}). The area values used for each centre computation are 
indicated in the upper panel of the figure, with the red and green boxes corresponding to the clusters 15408 and 15412, respectively. 
As a complement, we also indicate by a yellow box, the area corresponding to the field shown in Fig. \ref{rcw95_fig}.

For the determination of structural parameters, we applied a radial density profile (RDP) analysis using a function that describes 
the variation of the stellar number density ($stars/area$), with the distance from the center of the cluster defined by Equation 
\ref{3Pking} \citep{king1962,alves2012}:

\begin{equation}\label{3Pking}
 \sigma(r)=\sigma_0\left[    \frac{1}{\sqrt{1+ \bigg(  \frac{r}{r_c}  \bigg)^2 }}-\frac{1}{\sqrt{1+ \bigg(  \frac{r_t}{r_c}  \bigg)^2 }}    \right]^2+\sigma_f
\end{equation}
where $r_c$ is the core radius, $r_t$ is the tidal radius, $\sigma_0$ is the central star density and $\sigma_f$ is the field star 
density. The latter correspond to a minimum density level of the combined background and foreground stars, indicating the contribution 
of contaminating field stars per unit area throughout the region in the direction of the cluster. Furthermore, $\sigma_f$ is crucial 
for estimating the cluster limiting radius (hereafter cluster radius, $r_{cl}$), defined as the distance from the cluster centre to 
the point where the RDP reach the $\sigma_f$ level. Specifically, the algorithm first count the number of RDP points 
that fall below a given maximum tolerance interval above $\sigma_f$. When the number of these points reaches a fixed value, the point 
closest to the $\sigma_f$ is stored as the most probable radius for this iteration, which is repeated increasing the tolerance around 
$\sigma_f$. Finally, the estimation for $r_{cl}$ is obtained from the average of the set of radius values stored after each iteration, 
where its associated error correspond to the standard deviation of this average.

The RDP determination for both clusters was then performed using the ASteCA algorithm, which computes the RDP by generating square 
rings of increasing sides via an underlying 2D histogram (grid) in the spatial diagram of the analyzed region, counting the stars 
within each square ring and dividing by its area. In Fig. \ref{RDP} we show the RDP obtained for the clusters 15408 and 15412, where 
each dot represent the number of stars per unit area as described above, with the associated uncertainties represented by the error 
bars. The horizontal blue line indicates the $\sigma_f$ value, the red arrow indicates the determined $r_{cl}$ value and the gray 
region indicates the associated uncertainty. The 3-parameter King profile \citep{king1962} is indicated by the dashed green line and 
the core radius is represented by the vertical green line. The tidal radius of the cluster's King profile is indicated by the green 
arrow. In Table \ref{tabla_PDR} we list all parameter values obtained from the RDP analysis, where the different radii are 
shown in two rows, the upper one shows them in degrees and the lower one in parsecs. This result suggests that the two clusters found 
by \citet{dutra2003}, for which the estimation of centers and angular dimensions were based on visual inspection on the 2MASS Ks images, 
actually correspond to the single cluster 15408 in this work.

\begin{table*}
\centering
\caption{Structural parameters of both clusters 15408 and 15412.}
\label{tabla_PDR}
\resizebox{15cm}{!} {
\begin{threeparttable} 	
        {\small
  	\begin{tabular}{lcccc}
  	\hline
  	\hline
        ID Cluster & Centre                                      & Cluster Radius ($r_{cl}$) & Core Radius ($r_c$) & Tidal Radius ($r_t$) \\
                  &($ra\pm \sigma_{ra}$, $dec\pm \sigma_{dec}$)$_{J2000}$ & $(\degr)$         & $(\degr)$                   & $(\degr)$ \\
                  &							    & (pc)	     &  (pc) & (pc) \\
  	\hline  	
         15408     & ($236.179\pm 0.002$, $-54.11\pm 0.001$)     & $0.049\pm 0.003$ & $0.024\pm 0.003$ & $0.1\pm 0.03$ \\
		   &						   & $2.05\pm0.13$    & $1.01\pm0.13$    & $4.19\pm0.13$ \\
         15412     & ($236.247\pm 0.001$, $-54.143\pm 0.001$)    & $0.019\pm 0.002$ & $0.006\pm 0.002$ & $0.06\pm 0.06$ \\
		    &						  & $0.80\pm0.08$     & $0.25\pm0.08$    & $2.51\pm2.51$ \\
  	\hline
        \hline
	\end{tabular}
}
\end{threeparttable}
}
\end{table*}

\subsubsection{Luminosity and Mass Functions}

\begin{figure}
\includegraphics[width=0.95\linewidth]{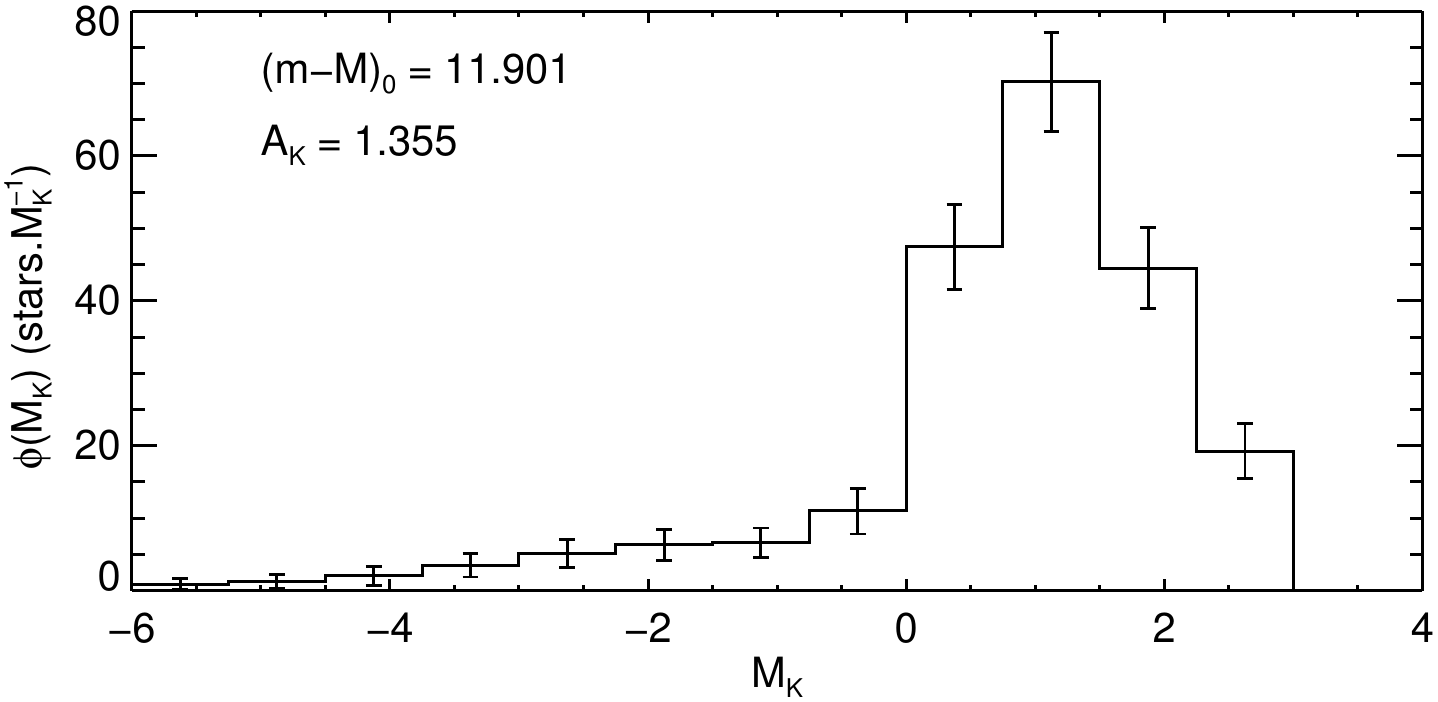}
\caption{M$_K$ luminosity function of the RCW95 region. Error bars were calculated using a Monte-Carlo procedure 
to propagate photometric and extinction uncertainties. The adopted distance modulus and extinction corrections are 
also shown.}
\label{fig:LF}
\end{figure}

The luminosity function (LF) of the RCW\,95 region was evaluated by adding together the identified members from both 
15408 and 15412 clusters. A $M_K$ LF was built by applying the distance modulus and reddening corrections to the stars 
magnitude and binning the sample using a $0.75$ magnitude interval. A Monte-Carlo procedure was used to propagate both 
the photometric uncertainties and the adopted extinction uncertainty into a LF uncertainty, by assuming a normal error 
distribution. The resulting LF is shown in Fig.~\ref{fig:LF}.

The mass function (MF) of RCW\,95 was then evaluated by using the mass-$M_K$ relationship from the 3.2 Myr 
($\log{t}=6.5$) PARSEC isochrone, the adopted age of both clusters. Mathematically, each bin $i$ of the LF, 
$\phi(M_{K,i}$), was converted into a mass function bin, $\xi(m_i)$, according to the following relation:

\begin{equation}
\xi(m_i) = \phi(M_{K,i})\frac{dM_K}{dm}\biggl\rvert_{M_{K,i}}
\end{equation}

\noindent
where $dM_K/dm$ represents the $M_K$-mass relationship, evaluated at the $M_{K,i}$ magnitude. Similarly, the 
$M_H$ and $M_J$ LFs were also used to obtain additional constraints to the mass function shape. Derivation of 
the total observed stellar mass ($\mathcal{M}$) is easily done through the following summation over all mass 
bins in a given band:

\begin{equation}
\mathcal{M} = \sum_i \xi(m_i)m_i
\end{equation}

\begin{figure}
\includegraphics[width=0.95\linewidth]{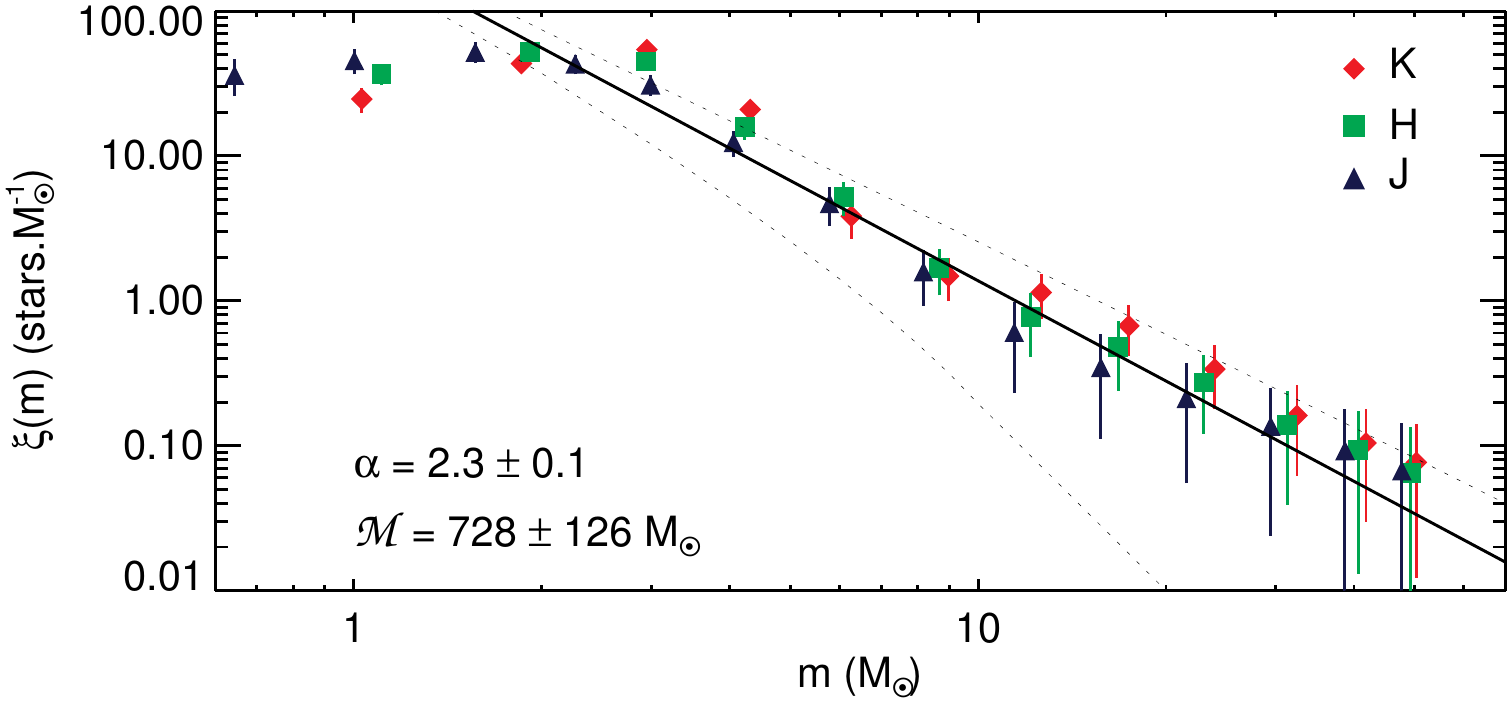}
\caption{Mass function of the RCW95 region, as derived from the $J$ (triangles), $H$ (squares) and $K_s$ (diamonds) 
bands. The total stellar mass observed and the power-law slope fitted to the higher-mass bins (solid line) are also shown.
Dotted lines show the region where stars do comply with the IMF within 1-sigma.}
\label{fig:mf}
\end{figure}

Fig.~\ref{fig:mf} shows the resulting MF for the RCW\,95 region, along with the derived total mass of observed 
stars. Since we are taking advantage of measurements in three bands, the total mass was obtained through an 
average of the summation in each band. The mass spectrum shows a power-law behavior at masses higher than 2 M$_\odot$, 
but flattens for lower masses in all filters. This result was also found in the mass function of each cluster 
individually, albeit at a lower confidence level, specially for cluster 15408. 

A fit of the stellar initial mass function (IMF), modelled by a power-law of the form $\xi(m) = A.m^{-\alpha}$ 
(hereafter K13 \citet{kroupa13}), was performed on the higher mass bins (m $\geq$ 2 M$_\odot$) yielding a slope of 
$\alpha=2.29 \pm 0.12$, in accordance with the expected IMF slope at this mass range. The flattening at lower mass 
bins probably signals higher than average incompleteness towards the clusters rather than selective mass loss driven 
by evolutionary effects. Although the cutoff limit at 2 M$_\odot$, corresponding to $K \sim 15.5$, is about 1 magnitude 
brighter than the derived completeness limit, crowding and higher extinction in young clusters is known to cause an 
increase in incompleteness towards its centre with relation to the mean incompleteness value derived from stellar 
counts \citep{m16}.

Given the youth of the region, it is unlikely that stars as massive as a solar mass star have been depleted by 
dynamical effects such as stellar evaporation. Therefore, the MF normalisation (calculated at 2 M$_\odot$) was 
used to extrapolate K13 IMF down to the hydrogen burning limit (0.08 M$_\odot$). Integration of this function 
to recover the unseen stellar content yielded a total stellar mass of $1220 \pm 213$ M$_\odot$ for the RCW95 
region, with 39\% held by the cluster 15408 and 61\% held by the cluster 15412. 

\subsubsection{Near-infrared spectroscopic classification of the most massive members and cluster ages}\label{CMD_section}

Through the use of CMDs built from the JHKs VVV photometry, combined with NIR follow-up spectroscopic observations of 
the main sources in each cluster, we were able to obtain valuable information about the stellar population of each cluster. 
The CMD in the direction of each cluster contain sources of both, background and foreground Galactic disk population. As 
a consequence the observed CMDs are the result of a mix of young and old stellar sources that in principle are difficult 
to interpret. In order to relieve this effect, we applied the decontamination method described in \citet{maia2010}, which 
is based on a statistical comparison between the observed CMD for each cluster and those in the direction of associated 
control regions. It enables the decontamination of the Galactic field population from the CMD of the cluster. The areas 
for the decontamination procedure were chosen accordingly with the results obtained in \S\ref{param_struct}, which are 
summarized in Table \ref{tabla_PDR}. The same control region was used for the decontamination of both clusters, 
which is centered at $\alpha_{\rm J2000}=15^{\rm h}44^{\rm m}38\fs6$ and $\delta_{J2000}=-54\degr18\arcmin53\arcsec$ (measuring 
about $2.8\times2.8$ square arcmin). It was selected from the visual inspection of the Spitzer 3.6\,$\mu$m and 8.0\,$\mu$m 
images of the region, looking for adjacent regions showing no extended emission at longer wavelengths, with the aim of avoiding 
as much as possible the areas immersed in the associated molecular complex. Finally, among all the possible control regions, we 
selected the nearest field with similar Galactic latitude $b$. 

\begin{figure*}
    \centering
  \includegraphics[width=0.8\textwidth]{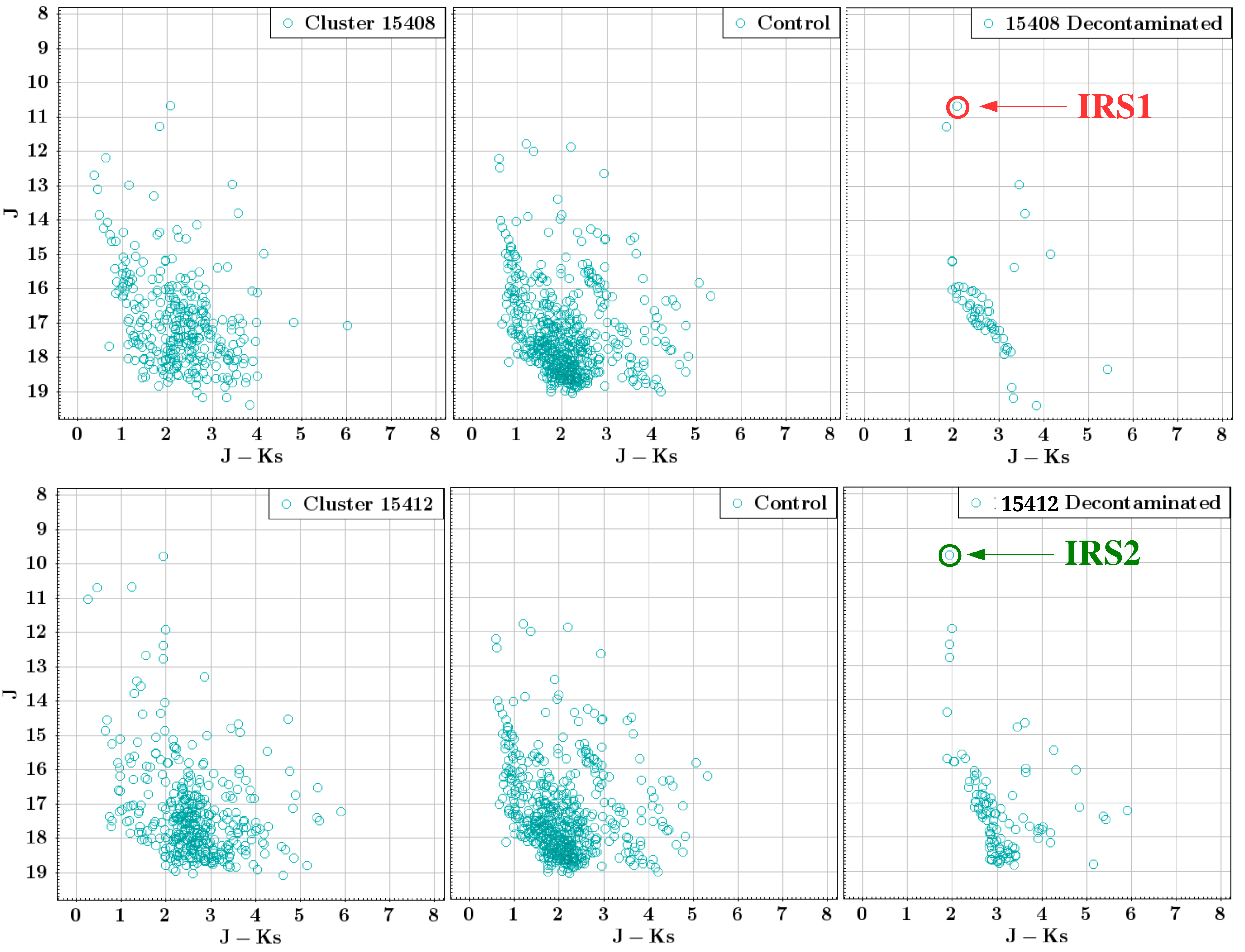}
   \caption{Colour-magnitude diagrams $J\times(J-Ks)$ of the clusters 15408 and 15412. ({\it left}) CMD in the direction of each cluster. 
           ({\it middle}) CMD inthe direction of the control region. ({\it right}) Decontaminated CMD of each cluster. There were 
           labeled those sources with near-infrared spectra (IRS1 and IRS2).}
\label{CMD_decon}
\end{figure*}

The resulting CMDs for each cluster are shown in Fig. \ref{CMD_decon} where the left panels correspond to the observed 
CMD in the direction of each cluster, the middle panel shows the CMD made from the sources found in the direction of the 
control region, while the decontaminated CMD for each cluster are represented by the right panels. We also indicate there 
the two sources for which we obtained NIR spectroscopic data that enabled us to refine the classification of the main 
ionizing source of the 15408 cluster, as well as to confirm the massive nature of the brightest source in the CMD of the 
15412 cluster. From an inspection of the two decontaminated CMDs, we can see that the associated vertical main-sequences 
are both seen at (J-Ks)$\sim$2.0 with the probable turn-on regions occurring approximately at J$\sim$16 mag, indicating 
that the two clusters are under about the same amount of interstellar extinction, and probably have similar ages.

\begin{figure*}
    \centering
 \includegraphics[width=0.8\textwidth]{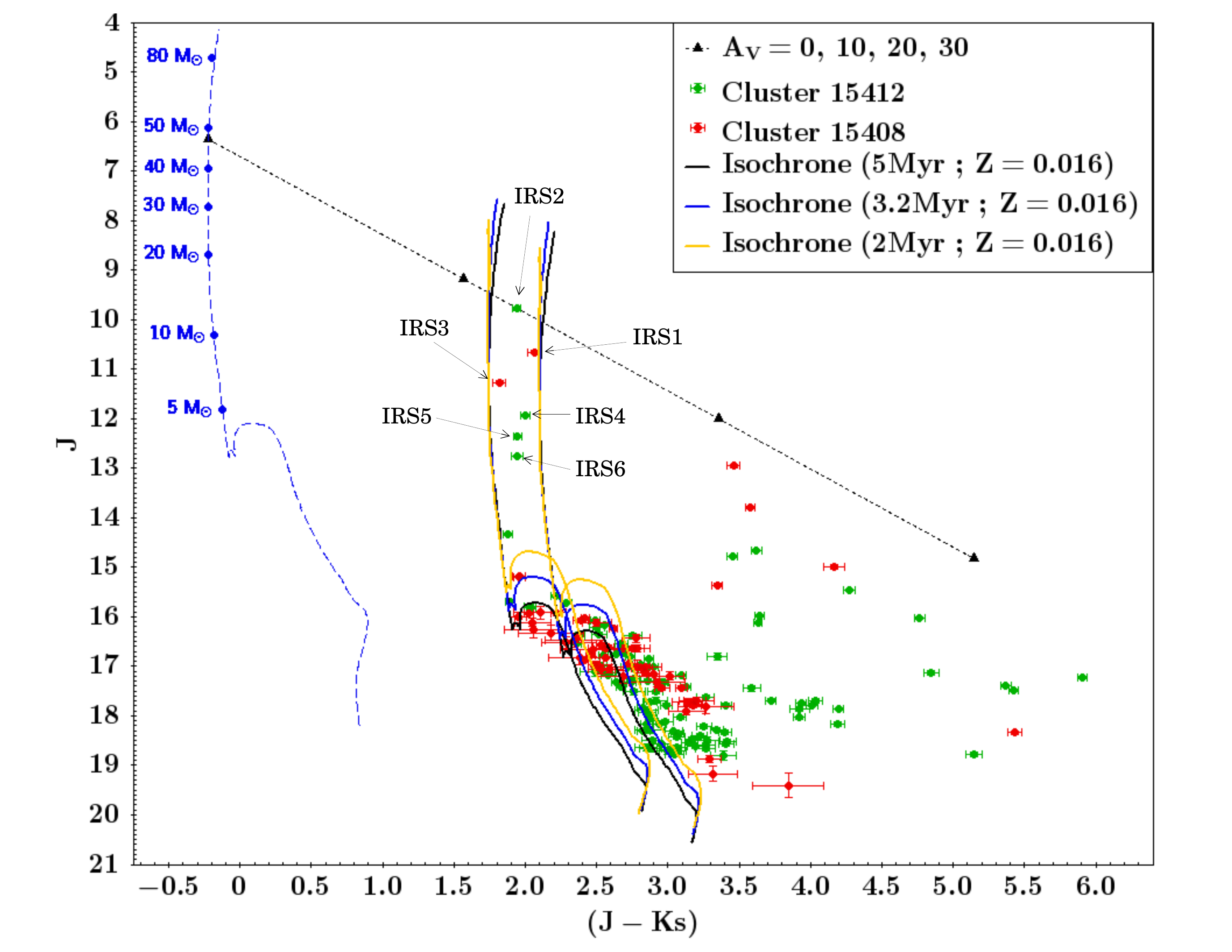}
  \caption{CMDs $J\times(J-Ks)$ of the decontaminated clusters 15408 (red) and 15412 (green). The yellow, blue and black solid lines correspond to a 2 Myr, 
           3.2 Myr and 5 Myr isochrones at $A_V=11\ -\ 13$ with Z=0.016 for a distance of 2.4 kpc. The dashed blue line correspond to the unreddened 3.2 Myr 
           isochrone where the respective stellar masses for MS stars are indicated. The dashed black line indicates the reddening vector for a O4 V star 
           \citep{martins2005} indicating the location of $A_V=$0, 10, 20 and 30 mag acording to interstellar extinction law from \citet{rieke1985} using 
           the $R_V=2.93\pm0.47$ from the polarimetric study. Stars marked with IRS1-6 labels are those in Table\ref{IRS1_6}.}
  \label{CMD_final}
\end{figure*}

Fig. \ref{CMD_final} shows the combined CMD containing point sources from the 15408 and 15412 clusters, which are represented 
by the red and green symbols, respectively. In this diagram we represent by the blue dashed line the non-reddened 3.2 Myr PARSEC
\citep{bressan2012,chen2015} isochrone (solar metallicity shifted to a heliocentric distance of 2.4 kpc), with each mass in the 
interval $5$ to $80\ M_{\odot}$ indicated. There we also show a set of PARSEC isochrones computed for ages 2.0 Myr, 3.2 Myr and 
5 Myr, reddened by $A_V=11$ and $13$.

%
%

\begin{figure*}
  \includegraphics[width=0.6\textwidth]{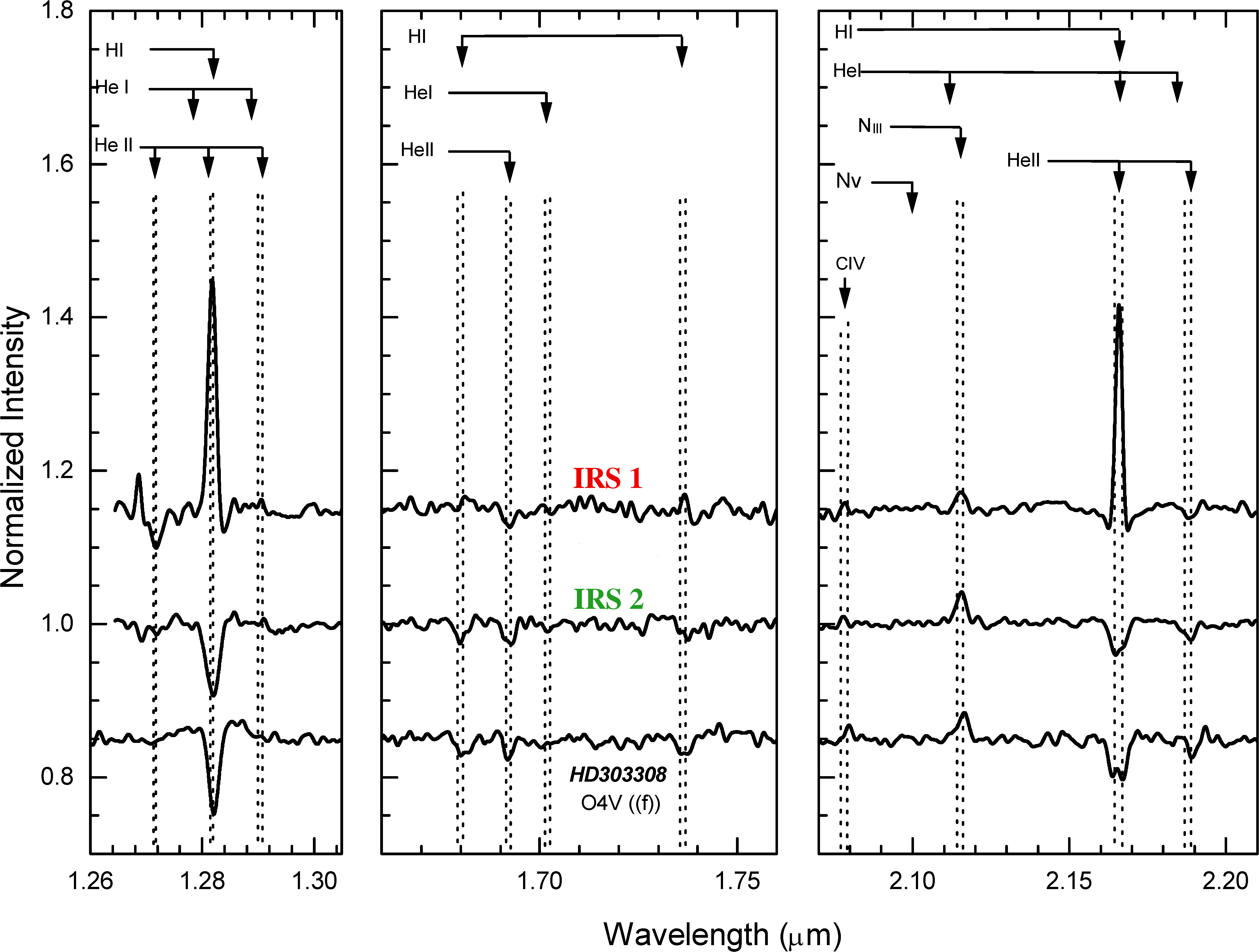}
  \caption{NIR spectra for the IRS 1 and IRS 2 sources in the J-, H- and K-bands. The reference spectra of the HD303308 star of 
           known spectral type O4V.}
  \label{spectra_fig}
\end{figure*}

From the position of sources IRS1 and IRS2 in the CMD of Fig. \ref{CMD_final}, and accordingly with the stellar models, we can see that 
both stars are probably very luminous objects with inferred masses of $40-50\ M_{\odot}$, which according to \citet{martins2005} should 
correspond to stars of O4V-O5V types. With our new SOAR-OSIRIS NIR spectra taken for the two brightest source in both clusters we confirmed
this assumption. The associated J-, H- and K-band spectra of IRS1 and IRS2 stars are presented in Fig. \ref{spectra_fig}. We also show the 
J-, H- and K-band template spectra of HD\,303308, a known O4V Galactic star. From the comparison of the former with the template spectra we 
can see that besides the residual hydrogen emission line features present in the IRS1 spectra, which could not be completely removed during 
the reduction process, all other $\textit{photospheric}$ lines seen in the J-, H- and K-bands are pretty similar, indicating that the three 
stars are probably of the same spectral type. Indeed, the J-, H-, K-band spectral features of IRS1 and IRS2 are very similar (besides the 
residual hydrogen emission line present in the spectra of the former) to those of HD\,303308 (O4V) and HD\,46223 (O4V) \citep{hanson2005}. 
The fact that the He{\sc i} $\lambda$17007 absorption lines are less intense than the He{\sc ii} $\lambda$16930 lines, and the presence of 
He{\sc ii} $\lambda\lambda$21890 absorption lines in the K-band, along with strong C{\sc iv} and N{\sc iii} $\lambda\lambda$20800-21160 
emission lines of similar intensity in all K-band spectra, led us to the conclusion that both sources are indeed early O-type stars probably 
of the O4V type. This spectral classification of the source IRS1 represents a more accurate classification than that previously determined by 
\citet{roman2009B} where they suggest a O5.5V type based in spectroscopic data in the K-band only, obtained with the Gemini Near-Infrared 
Spectrograph (GNIRS) in Gemini South. Our new classification incorporates the analysis of J- and H-bands which include important spectral 
lines of hydrogen and helium for a further discrimination between early-O type stars.

\citet{Bik2005} performed a K-band spectral classification of IRAS point sources that were selected based on their position in the 
colour-magnitude diagram, where the source 15408nr1410, presenting $(J-Ks)=2.2\pm0.02$ and $Ks=8.6\pm0.01$ in their work, is classified as a O5V-O6.5V 
star. This source actually corresponds to the IRS1 source in this work, also presented as the ionizing source in the center cluster.

In order to derive the mean visual extinction in the direction of the clusters we used the intrinsic colours \citep{martins2006} of their 
brightest sources (marked with IRS1-6 labels in Fig. \ref{CMD_final}) and the interstellar extinction law given by \citet{rieke1985}. 
In this way, it is possible to derive the colour excesses $E(J-H)$ and $E(B-V)$ through $E(J-H)/E(B-V)=0.33\pm0.04$. Finally, by using 
the total-to-selective extinction ratio $R_V$ derived in this work and the relation $R_V=A_V/E(B-V)$ we obtained a visual extinction 
value for each source and a mean visual extinction of $A_V=12.1\pm6.0$. Although the uncertainties of the $E(B-V)$ range around $~12\%$, 
the resulting uncertainty of the mean value of $A_V$ reaches up to $50\%$, this arise from the high dispersion of $R_V$ values. This result 
is summarized in Table \ref{IRS1_6}.

\begin{table*}
\centering
\caption{Summary of extinction values of sources labeles in Fig. \ref{CMD_final}.}
\label{IRS1_6}
\resizebox{17cm}{!} {
\begin{threeparttable} 	
        {\small
  	\begin{tabular}{lcccccccc}
 	\hline
 	\hline
        ID	& $\alpha_{2000} (\ ^{\rm h}\ ^{\rm m}\ ^{\rm s})$& $\delta_{2000}( \degr \ \arcmin \ \arcsec )$ & $(J-H)_o$ & $E(J-H)$ & $E(B-V)$ & $A_V$    & $A_K$   & Spectral Type\\
        	& 					 &					& 	    &		&	   & \multicolumn{2}{c}{($R_V=2.93\pm0.47$)}&  \\
 	\hline  	
        IRS1	&15:44:43.39	&-54:05:53.5	&-0.11	&$1.342\pm0.051$ &$4.067\pm0.470$ &$11.920\pm2.438$& $1.335\pm2.438$ & O4 V \\
        IRS2	&15:44:59.55	&-54:08:47.6  	&-0.11	&$1.356\pm0.037$ &$4.109\pm0.510$ &$12.039\pm2.442$& $1.348\pm2.442$ & O4 V\\	 
        IRS3	&15:44:39.74	&-54:06:32.0	&-0.11	&$1.206\pm0.051$ &$3.655\pm0.469$ &$10.709\pm2.199$& $1.199\pm2.199$ &\\	 
        IRS4	&15:44:57.91	&-54:08:21.5  	&-0.11	&$1.473\pm0.037$ &$4.464\pm0.553$ &$13.080\pm2.649$& $1.465\pm2.649$ &\\ 
        IRS5	&15:44:55.80	&-54:08:10.7  	&-0.11	&$1.408\pm0.031$ &$4.267\pm0.526$ &$12.502\pm2.528$& $1.400\pm2.528$ &\\	 
        IRS6	&15:45:03.96	&-54:08:08.2  	&-0.11	&$1.376\pm0.048$ &$4.170\pm0.526$ &$12.218\pm2.493$& $1.368\pm2.493$ &\\	
 	\hline
        \hline
 	\end{tabular}  
}
\end{threeparttable}
}
\end{table*}

\section{Conclusions}

In this work we performed a study of the interstellar linear polarization component in the direction of the RCW\,95 Galactic 
H{\sc ii} region, a massive, dense and rich star-forming region. The behaviour of the magnetic field lines that permeate
this star-forming region was determined, and from the empirical Serkowski relation \citep{b16} applied to the polarimetric 
data, we were able to compute the total-to-selective extinction ratio ($R_V$) mean value in this direction of the Galaxy. Also, 
another objective of this work was to improve the study of the clusters stellar population, which was done with PSF photometry
applied to new near-infrared VVV images, combined with J-, H- and K-band spectroscopic data taken with OSIRIS at SOAR.

From the analysis of the spectroscopic data and related optical-NIR photometry, our main results are:

\begin{itemize}

\item The CTIO optical V-, R- and I-band polarimetric maps look similar in all three spectral bands, with the mean values 
$\langle\theta_V\rangle=51\fdg2\pm7\fdg6$, $\langle\theta_R\rangle=49\fdg8\pm7\fdg7$ and $\langle\theta_I\rangle=49\fdg4\pm8\fdg9$ 
being compatible considering the associated uncertainties. The overall polarization segments distribution seems to be well 
aligned with the more extended cloud component, with a mean polarization angle of $49\fdg8$.

\item The polarization degree values for the best quality data ranges between $P=0\ -\ 10\%$ with the higher 
frequencies occurring for values between $P=3\ -\ 6\%$ (with $\sim 60\%$ of the source distribution lying within 
this range). For the assumed RCW\,95 heliocentric distance of 2.4 kpc, it is expected that the polarization 
levels measured for the stars in our sample are affected by a line of sight foreground interstellar component. 
In order to estimate this mean foreground component, we made use of the VVV NIR photometric colours of the 
probable foreground and background sources, and based on the observed connection with the associated polarimetric 
parameters we estimated the line of sight foreground interstellar component of the polarization degree in the 
direction of RCW95 as $P_R=2.0\%$.

\item Based on the polarimetric dataset derived from the V, R and I bands observations, it was possible to study the 
wavelength dependence of the polarization degree in the direction of RCW\,95, which in turn enabled us to compute a 
mean value of the total-to-selective extinction ratio $R_V=2.93\pm0.47$, which combined with the measured colour excess 
derived from the OB sources found in both clusters resulted in a mean visual absorption of A$_V$=12.1$\pm6.0$ magnitudes.

\item Using the Automated Stellar Cluster Analysis (ASteCA) algorithm developed by \citet{perren2015}, and the statistical 
decontamination method of \citet{maia2010}, we derived estimates for the cluster limiting radius $r_{cl}$, core radius 
$r_{core}$ and tidal radius $r_{t}$ of the two clusters in the RCW\,95 region. From the photometry data coupled with a set 
of PAdova and TRieste Stellar Evolution Code (PARSEC) isochrones \citep{bressan2012,chen2015}, we estimated an age of about 
3 Myrs for both clusters.

\item The LF of the RCW\,95 region was evaluated from the identified members of both 15408 and 15412 clusters and its MF 
determined from the mass-$M_K$ relationship of the 3.2 Myr ($\log{t}=6.5$) PARSEC isochrone, the adopted age of both clusters.
The $M_J$ and $M_H$ LFs were also considered to obtain additional constraints on the MF shape. The resulting mass spectrum 
shows a power-law behaviour at masses higher than $2 M_{\odot}$ but flattens for lower masses in all filters. The stellar 
IMF was evaluated for m $\geq$ 2 M$_\odot$, yielding a slope of $\alpha=2.29\pm0.12$, in accordance with the expected IMF 
slope at this mass range. MF normalisation was used to extrapolate the K13 IMF down to the hydrogen burning limit (0.08 
M$_\odot$) that finally yields a total stellar mass of 1220$\pm$213 M$_\odot$ for the RCW\,95 region, shared by the clusters 
15408 (39\%) and 15412 (61\%).

\item These results, together with the compact nature of the clusters as derived from their structural parameters, are in 
agreement with star formation scenarios in which a number of small sub-clusters form first from a molecular cloud and then 
merge to produce fewer larger ones \citep{krumholz2014}.

\item From our near-infrared photometric study of VVV images of the RCW95 region, combined with new SOAR-OSIRIS NIR spectra 
of the two brightest sources of the sample, we were able to refine the spectral classification of the main ionizing source of 
the 15408 cluster, as well as identifying the existence of another early-O star, unknown until the present and the dominant 
source of Lyman photons in the new cluster 15412 identified there. Both objects are found to be O4V stars, which are the 
main ionizing sources of the H{\sc ii} region.
\end{itemize}

\section*{Acknowledgments}

This work is based [in part] on observations made with the Spitzer Space Telescope, which is operated by the Jet Propulsion 
Laboratory, California Institute of Technology under a contract with NASA. This research made use of data obtained at the 
Southern Astrophysical Research (SOAR) telescope, which is a joint project of the Minist\'{e}rio da Ci\^{e}ncia, Tecnologia, 
e Inova\c{c}\~{a}o (MCTI) da Rep\'{u}blica Federativa do Brasil, the U.S. National Optical Astronomy Observatory (NOAO), the 
University of North Carolina at Chapel Hill (UNC), and Michigan State University (MSU). We acknowledge the use of 
data from the ESO Public Survey program ID 179.B-2002 taken with the VISTA 4.1 m telescope. JV-G acknowledges the financial 
support of the Direcci\'{o}n de Investigaci\'{o}n of the Universidad de La Serena (DIULS), through a ``Concurso de Apoyo a 
Tesis 2013'', under contract N$^o$ PT13147. G.A.P.F. acknowledges the partial support from CNPq and FAPEMIG (Brazil).

\bibliography{references}

\label{lastpage}

\end{document}